# From Multi-Keyholes to Measure of Correlation and Power Imbalance in MIMO Channels: Outage Capacity Analysis

Georgy Levin and Sergey Loyka

*Abstract*—An information-theoretic analysis of a multi-keyhole channel, which includes a number of statistically independent keyholes with possibly different correlation matrices, is given. When the number of keyholes or/and the number of Tx/Rx antennas is large, there is an equivalent Rayleigh-fading channel such that the outage capacities of both channels are asymptotically equal. In the case of a large number of antennas and for a broad class of fading distributions, the instantaneous capacity is shown to be asymptotically Gaussian in distribution, and compact, closed-form expressions for the mean and variance are given. Motivated by the asymptotic analysis, a simple, full-ordering scalar measure of spatial correlation and power imbalance in MIMO channels is introduced, which quantifies the negative impact of these two factors on the outage capacity in a simple and well-tractable way. It does not require the eigenvalue decomposition, and has the full-ordering property. The size-asymptotic results are used to prove Telatar's conjecture for semi-correlated multi-keyhole and Rayleigh channels. Since the keyhole channel model approximates well the relay channel in the amplify-and-forward mode in certain scenarios, these results also apply to the latter.

## I. INTRODUCTION

OUTAGE capacity is one of the major characteristics of fading channels, as it provides an ultimate upper limit on the error-free information rate with a given probability of outage [1][2]. The outage capacity of spatially independent as well as correlated Rayleigh, Rice and Nakagami MIMO channels has been extensively studied, and a number of analytical and empirical results have been obtained [1]-[9]. There are, however, propagation environments that result in substantially different channels. Chizhik *et al* [10][11] have introduced a keyhole channel as a worst-case MIMO propagation environment. This channel can be represented as a cascade of two Rayleigh-fading channels separated by a keyhole whose dimensions are much smaller than the wavelength. The presence of the keyhole degenerates the channel, i.e. its rank is one regardless of the number of Tx and Rx antennas. Consequently, the techniques exploiting the multiplexing gain of MIMO channels to increase the data rate (e.g. BLAST) become inefficient. In contrast, the methods taking advantage of the spatial diversity, such as Alamouti scheme [12], are beneficial and substantially reduce the error rate. The interest in the keyhole channels has recently increased since they appear in some practically important propagation scenarios. A keyhole scenario where the propagation path between Tx and Rx ends is due to the 1-D edge diffraction is given in [11]. The outdoor model in [13] suggests the existence of the keyhole channel in a rich scattering environment, where the scattering rings around the Tx and Rx antennas are small comparing to the distance between the Tx and Rx ends. Using empirically validated channel model, [14] shows that the mean capacity of a free space propagation channel follows closely the corresponding capacity of the keyhole channel, when the distance between the Tx and Rx ends is large. In [15], the indoor measurement taken along hallways exhibited a decrease in the channel capacity with distance, which is explained by the keyhole effect. The first convincing experimental evidences of the keyhole channel has been demonstrated in laboratory environment in [16][17], where it was shown that the keyhole model describes wireless channels when the wave propagates via waveguides. The waveguide channel with a single propagating mode, which can model certain indoor scenarios [18], is also an example of a keyhole channel.

There are a number of studies that provide information theoretic analysis of the keyhole channels. The mean and outage capacities of spatially uncorrelated and correlated keyhole channels have been studied in [21]-[23]. The diversity order of uncorrelated keyhole channels is investigated in [24][25]. The performance analysis of space time block codes over uncorrelated keyhole Nakagami-m fading channels has been carried out in [26].

However, keyhole channels are not often encountered in practice, since the assumption of a single propagation eigenmode is only a rough approximation of real propagation environments. It has been shown in [16] that the keyhole effect is difficult to observe, since, in many scenarios, the contribution of other eigenmodes cannot be neglected. To include these scenarios and expand the application range of the keyhole channel model, a multi-keyhole channel, which includes a number of statistically independent keyholes, was introduced in [19]-[21].

The multi-keyhole channel is closely related to the double-scattering model in [13], since the keyholes and scatterers perform essentially the same function of re-radiating the transmitted signal. The significant difference between these two models is due to the fact that they represent different geometric configurations and assume different fading statistics:

- The double-scattering model in [13] assumes that sub-channels corresponding to different scatterers (equivalently, keyholes) are described by the same correlation matrix, and are also correlated with each other. Since the eigenvectors of a correlation matrix correspond to the steering vectors at the directions of energy transmission/reception [27], this implies that the scatterers are located close to each other, which is also consistent with the fact that different sub-channels are correlated with each other. Thus, the double-scattering model represents dense scattering environment.

- The multi-keyhole model considered here assumes that sub-channels associated with different keyholes are described by different correlation matrices (including the special case when they are equal) and that they are independent of each other, which corresponds to a sparse scattering environment (when the keyholes are far away from each other).

- Contrary to [13], no specific assumptions about the channel fading distribution (e.g. Rayleigh fading) are made for the multi-keyhole model (only Theorem 1 requires such an assumption; other results hold for a broad class of fading distributions).

Thus, these two models are essentially complementary to each other.

A detailed information-theoretic analysis of the multi-keyhole channel model is not available yet[1]. While Gesbert *et al* [13] introduced the double-scattering model, its channel capacity was evaluated via simulations only, without underlying information-theoretic analysis. The outage capacity of multi-keyhole or double-scattering channels has not been found and the impact of correlation, number of keyholes/scatterers and other parameters has not been studied. No comparative analysis between the multi-keyhole/double-scattering and canonic channels, such as Rayleigh-fading, has yet been made. This paper fills these gaps by providing new results on the outage capacity of multi-keyhole channels correlated at both ends. While the exact outage capacity of MIMO channels is rarely amenable to a closed-form analysis, we obtain compact closed-form approximations via the asymptotic analysis with respect to the number of antennas/keyholes, which results in a number of important insights and applications. This asymptotic approach has already been successfully applied to Rayleigh-fading [30]-[32] and single-keyhole MIMO channels [21], and has been found to predict reasonably well the performance when the number of antennas is moderate, and it is extended here to multi-keyhole channels. In particular, we show that the instantaneous capacity[2] of multi-keyhole channels is asymptotically Gaussian, under mild assumptions and for a broad class of fading distributions and correlation models. This along with other asymptotic results in the literature (e.g. in [33][38][3]) suggests that the Gaussian distribution has a high degree of universality for outage capacity analysis of MIMO channels in general.

Based on the asymptotic analysis of a single-keyhole channel, a scalar measure that characterizes the impact of correlation on the outage capacity has been introduced in [19]-[21]. This measure has also been shown to characterize the impact of correlation on the mean capacity and diversity gain in Rayleigh-fading and double-scattering channels [34][35][45]. In this paper, a similar measure is shown to characterize the effects of correlation and also power imbalance on the outage capacity of multi-keyhole channels for a broad class of fading distributions. The introduced measure is shown to be always finite (even asymptotically, when the number of antennas/keyholes increases to infinity), it does not require eigenvalue decomposition (i.e. simple to evaluate), unlike the measures based on the majorization theory [37], it has full ordering property (any two channels can be compared, without exceptions), it clearly separates the effect of correlation and power imbalance (none of the known measures do), and it is compatible with the corresponding measure based on the majorization theory [37].

The size-asymptotic approach opens a possibility to attack a number of problems, which are associated with significant mathematical complexity when the number of antennas is finite. While Telatar's conjecture [1] has been proven only for MISO and SIMO Rayleigh-fading channels [36] and remains an open problem in general, we provide a compact proof of the conjecture asymptotically for semi-correlated multi-keyhole (for a broad class of fading distributions) and Rayleigh-fading channels.

It should be noted that the keyhole channel can also serve as a model of the relay channel in the amplify-and-forward mode (see [28] for related models and results on relay channels). Specifically, in many practically-important cases the relay noise can be neglected (see [61] for details), and therefore, the relay channel in these cases is well approximated by a keyhole one, where the keyhole represents a relay node rather than a propagation mechanism. This opens up a possibility to apply the keyhole channel results to relay channels as well. In particular, we note that the throughput gain from transmission scheduling in a multiuser environment with "amplify-and-forward" relays and the feedback rate required to support that throughput can be estimated using the approach developed in [30] for asymptotically large Rayleigh-fading channels, and the estimates are valid for a broad class of fading distributions, not only Rayleigh one.

The main results of the paper are summarized as follows:

- The instantaneous capacity of a multi-keyhole channel is either upper bounded (finite number of keyholes) by or converges in distribution (number of keyholes increases to infinity) to that of an equivalent Rayleigh-fading channel

---

[1] After this paper had been submitted, capacity and outage analysis of beamforming in multi-keyhole and double-scattering channels have been presented in [58]-[60].

[2] i.e. the capacity of a given channel realization.

[3] Although [[33], Theorem 2.76] refers to a wide range of MIMO channels including a Rayleigh one, it cannot be applied to keyhole and multi-keyhole channels since these channels are not spatially independent, and, furthermore, they cannot be generated by a linear combination of independent components, e.g. as in [38].

(Theorem 2, Corollaries 2.1-2.3).

• Likewise, the instantaneous capacity of a multi-keyhole channel is either upper bounded (finite number of antennas) by or equals in probability (number of Tx or Rx antennas increases to infinity) to that of an Rayleigh-fading channel (Theorems 1, 3).

• While the instantaneous capacity of full-rank and rank-deficient multi-keyhole channels is asymptotically (large number of antennas) Gaussian, they have different means/variances and the effect of correlation is different (Theorems 3, 4).

• An additional motivation for the Kronecker correlation model (see [29] for details on this model) is provided by considering the Rayleigh-fading channel as a multi-keyhole one with a large number of keyholes.

• Measures of correlation and power imbalance that clearly separate these two effects and represent adequately the outage capacity of multi-keyhole channels are introduced and investigated (Sections IV, V).

• Telatar's conjecture [1] is proved for semi-correlated multi-keyhole channels (for a broad class of fading distributions) and also generalized to semi-correlated Rayleigh channels, and the set of active antennas is identified (Theorem 6, Corollary 6.1).

The paper is organized as follows. The system model is introduced in Section II. The outage probability and capacity of multi-keyhole channels is analyzed in Section III. Section IV studies the measure of channel correlation and power imbalance. The impact of correlation and power imbalance on outage capacity is discussed in Section V. Telatar's conjecture is considered in Sections VI. Section VII concludes the paper. Proofs are given in the Appendix.

## II. MULTI-KEYHOLE CHANNEL MODEL

The following discrete-time, baseband model of a MIMO channel with $n_t$ Tx and $n_r$ Rx antennas is used,

$$\mathbf{y} = \mathbf{H}\mathbf{x} + \mathbf{w}, \qquad (1)$$

where $\mathbf{x}$ and $\mathbf{y}$ are transmit and receive vectors respectively, $\mathbf{H}$ is the channel matrix whose elements $H_{km}$, $k = 1...n_r$, $m = 1...n_t$, represent the complex channel gains from $m$-th transmit to $k$-th receive antennas, and $\mathbf{w}$ is the AWGN noise vector. Unless otherwise indicated, we adopt the following assumptions:

(i) the channel state information (CSI) is available at the Rx end only,

(ii) $\mathbf{x} \propto \mathcal{CN}(\mathbf{0}, P_T/n_t \mathbf{I})$, where $\propto$ means identically distributed, $\mathbf{I}$ is identity matrix, and $P_T$ is the total transmitted power, which does not depends on $n_t$ (this achieves the ergodic capacity of the i.i.d. Rayleigh fading channel [1], the outage capacity under certain conditions (see e.g. [36][49], Theorem 6 and Corollary 6.1 in the present paper) and is a reasonable transmission strategy with no Tx CSI in general [62][66]),

(iii) the total average received power $P_R$ is constant regardless of $n_r$ (this corresponds to a densely-populated antenna array [39] of fixed apperture),

(iv) $\mathbf{w} \propto \mathcal{CN}(\mathbf{0}, N_0 \mathbf{I})$, where $N_0$ is the noise variance in each receiver,

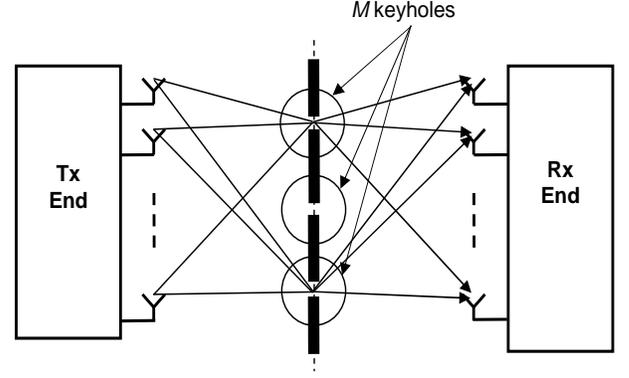

Fig. 1. A multi-keyhole MIMO channel model: each end has its own set of multipath components and is separated from the other end by a screen with a number of keyholes of small size (smaller than half a wavelength).

(v) the channel is frequency flat and quasi-static (slow block fading).

Following [2], the instantaneous capacity of such a MIMO channel in natural units $[nat/sec/Hz]$ is given by

$$C = \ln \det \left( \mathbf{I} + \frac{\gamma_0}{n_t n_r} \mathbf{H}\mathbf{H}^\dagger \right), \qquad (2)$$

where $\det$ denotes a determinant, $\mathbf{H}^\dagger$ is the Hermitian transpose of $\mathbf{H}$, and $\gamma_0 = P_R/N_0$ is the total SNR at the Rx end. $\mathbf{H}$ is normalized so that $E \|\mathbf{H}\|^2 = \sum_{i,j} E |H_{i,j}|^2 = n_t n_r$, where $\|\cdot\|$ is the Frobenius norm.

Consider a spatially correlated multi-keyhole MIMO channel (see Fig. 1). The channel matrix $\mathbf{H}$ is given by the following linear combination [21]:

$$\mathbf{H} = \sum_{k=1}^{M} a_{k,M} \mathbf{h}_{rk} \mathbf{h}_{tk}^\dagger = \mathbf{H}_r \mathbf{A} \mathbf{H}_t^\dagger, \qquad (3)$$

where $M$ is a number of keyholes, $a_{k,M}$ is the complex gain of $k$-th keyhole, $\mathbf{h}_{tk}$ and $\mathbf{h}_{rk}$ are random $n_t \times 1$ and $n_r \times 1$ vectors representing the complex gains from the transmit antennas to the $k$-th keyhole and from the $k$-th keyhole to the receive antennas respectively; $\mathbf{H}_t = [\mathbf{h}_{t1}..\mathbf{h}_{tM}]$, $\mathbf{H}_r = [\mathbf{h}_{r1}..\mathbf{h}_{rM}]$, and $\mathbf{A} = diag\{a_{k,M}\}$ is a diagonal matrix. Assume that

(i) the keyholes are statistically independent, i.e.

$$E\{\mathbf{h}_{tk}\mathbf{h}_{tm}^\dagger\} = E\{\mathbf{h}_{rk}\mathbf{h}_{rm}^\dagger\} = \mathbf{0} \quad \forall k \neq m,$$

(ii) $\mathbf{h}_{tk}$ and $\mathbf{h}_{rk}$ are normalized, so that for every $k$,

$$n_t^{-1} E \|\mathbf{h}_{tk}\|^2 = n_r^{-1} E \|\mathbf{h}_{rk}\|^2 = 1,$$

which implies, under $E \|\mathbf{H}\|^2 = n_t n_r$, that

$$\sum_{k=1}^{M} |a_{k,M}|^2 = 1, \qquad (4)$$

Since the average power at the Rx end is proportional to $E \|\mathbf{H}\|^2$ [39], normalization (4) implies that $P_R$ does not depend on the number of keyholes, i.e. the total "cross section" of the channel is assumed to be constant. Note that, contrary to [13][22], no specific assumptions (e.g. Rayleigh fading) about the distribution of $\mathbf{H}$ are made at this stage.

While the multi-keyhole model above has a structure similar to that of the double-scattering model in [13], there are a number of essential differences, as discussed in the Introduction.

Note that the two models are identical when the sub-channels of different keyholes are independent ($\mathbf{h}_{r(t)i}$ is independent of $\mathbf{h}_{r(t)j}$, $i \neq j$) and when these sub-channels have the same correlation matrix, so that our results apply to the double-scattering model as well in that case.

## III. Capacity and Outage Probability of Multi-Keyhole Channels

In this section, we study the capacity distribution of the multi-keyhole channel and its relationship to the canonical Rayleigh-fading channel. The instantaneous capacity of the multi-keyhole channel is [21]

$$C = \ln \det \left( \mathbf{I} + \gamma_0 \mathbf{B}_r \mathbf{A} \mathbf{B}_t \mathbf{A}^\dagger \right), \quad (5)$$

where $\mathbf{B}_t = \mathbf{H}_t^\dagger \mathbf{H}_t / n_t$ and $\mathbf{B}_r = \mathbf{H}_r^\dagger \mathbf{H}_r / n_r$.

The outage probability is defined as the probability that the channel is not able to support target rate $R$, i.e. $P_{out} = \Pr\{C < R\}$, and the corresponding outage capacity is defined as the largest possible rate such that the outage probability does not exceed the target value $\epsilon$ [62],

$$C_\epsilon = \max\{R : P_{out}(R) \leq \epsilon\} \quad (6)$$

When $P_{out}(R)$ is monotonically increasing in $R$, $C_\epsilon = P_{out}^{-1}(\epsilon)$, where $P_{out}^{-1}$ denotes the functional inverse of $P_{out}(R)$, so that $P_{out}(C_\epsilon) = \epsilon$. Following Root and Varaya's compound channel capacity theorem [63], $C_\epsilon$ is achievable by a single universal code of a rate arbitrary close to $C_\epsilon$ on any channel that is not in the outage set (see also [64][65] for a modern approach). Likewise, such a code also achieves the block error rate equal to the channel outage probability (for a given target rate). An alternative interpretation of (6) is via an adaptive-rate system: the transmitter knows the instantaneous channel capacity $C$ and sets the transmission rate arbitrary close to it, which achieves simultaneously the instantaneous capacity $C$, the outage probability $P_{out}(R)$ (for given target rate $R$) or the outage capacity $C_\epsilon$ (for target outage probability $\epsilon$).

The following theorem indicates the relationship between the multi-keyhole and the canonic Rayleigh-fading channels.

*Theorem 1:* (i) Consider a multi-keyhole channel with $M$ independent keyholes, such that $\mathbf{h}_{tk}$ and $\mathbf{h}_{rk}$ are mutually independent complex circular symmetric Gaussian vectors with corresponding correlation matrices $\mathbf{R}_{tk} = E\{\mathbf{h}_{tk} \mathbf{h}_{tk}^\dagger\}$, $\mathbf{R}_{rk} = E\{\mathbf{h}_{rk} \mathbf{h}_{rk}^\dagger\}$, and

$$\lim_{n_t \to \infty} n_t^{-1} \|\mathbf{R}_{tk}\| = 0, \quad k = 1...M.$$

Then, as $n_t \to \infty$, there exists an equivalent Rayleigh-fading channel, such that the instantaneous capacities of both channels are equal in probability, i.e.

$$C \xrightarrow{p} \ln \det \left( \mathbf{I} + \frac{\gamma_0}{n_r} \mathbf{H}_r \mathbf{Q} \mathbf{H}_r^\dagger \right), \quad (7)$$

where $\xrightarrow{p}$ denotes convergence in probability, $\mathbf{H}_r$ represent the equivalent Rayleigh-fading channel, and $\mathbf{Q} = \mathbf{A}\mathbf{A}^\dagger$ is the diagonal power allocation matrix in the equivalent channel. It follows that the corresponding outage probabilities/capacities are also equal,

$$\Pr\{C < R\} = \Pr\left\{ \ln \det \left( \mathbf{I} + \frac{\gamma_0}{n_r} \mathbf{H}_r \mathbf{Q} \mathbf{H}_r^\dagger \right) < R \right\}, \quad (8)$$

where $R$ is the target rate (or the outage capacity for a given outage probability).

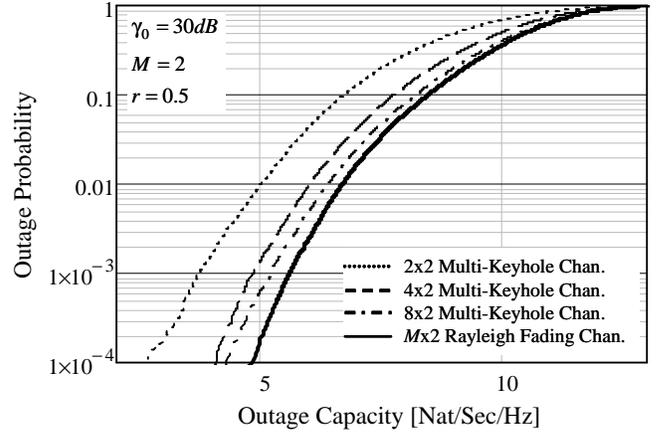

Fig. 2. Outage probability/capacity of multi-keyhole correlated Rayleigh-fading channels.

(ii) Due to the symmetry in (5), this also holds true when Tx and Rx ends are exchanged as $n_r \to \infty$.

*Proof:* see Appendix. ∎

The following arguments give intuition behind Theorem 1. For large $n_t$, the Tx sub-channel (see Fig. 1) is asymptotically non-fading AWGN due to the large diversity order ($= n_t$), so that the end-to-end channel becomes Rayleigh-fading with $M$ Tx antennas (i.e. keyholes), each with the power gain $|a_{k,M}|^2$. Similarly, when $n_r$ is large, the end-to-end channel is asymptotically Rayleigh-fading with $M$ Rx antennas, each with the power gain $|a_{k,M}|^2$. Theorem 1 generalizes the corresponding result in [21] obtained for $M = 1$.

As an example, consider the capacity distributions of $n_t \times 2$ multi-keyhole channels with two independent keyholes ($|a_{1,2}| = |a_{2,2}| = 1/\sqrt{2}$) and the equivalent 2x2 Rayleigh-fading channel shown in Fig. 2. The Kronecker model [29] is used to simulate the correlation in the Rayleigh-fading channel. The correlation matrices for both the multi-keyhole and Rayleigh-fading channels are modeled using the exponential correlation model [40] with the adjacent antenna correlation $r = 0.5$ at both ends[4]. From Fig. 2, the outage capacity increases with the number of Tx antennas and asymptotically approaches the capacity of the equivalent 2x2 Rayleigh-fading channel (a bold solid line). In the considered range of outage probabilities, the difference between the two becomes practically negligible for $n_t = 8$.

Note that given the same outage probability and $n_r$, the outage capacity of the equivalent Rayleigh-fading channels is always higher than that of the multi-keyhole one. The Rayleigh channel capacity is achieved only asymptotically as $n_t \to \infty$.

---

[4] Unless otherwise is indicated, the Kronecker and exponential correlation models are applied throughout the paper in examples.

In the following discussion we distinguish between two different types of multi-keyhole channels:

(i) a full-rank multi-keyhole (FRMK) channel, where $M \geq \min\{n_t, n_r\}$,

(ii) a rank-deficient multi-keyhole (RDMK) channel, where $M < \min\{n_t, n_r\}$.

It is straightforward to show that similarly to the Rayleigh-fading channel, the multiplexing gain [41] of the FRMK channel is limited by $\min\{n_t, n_r\}$. In contrast, that of the RDMK channel is limited by $M$. Below we show that given the same $\gamma_0$, FRMK and RDMK channels have different outage probabilities/capacities, and the impact of correlation is via different mechanisms.

### A. Full-Rank Multi-Keyhole Channel

Below we show that the FRMK channel is asymptotically Rayleigh-fading as $M \to \infty$. While this result is intuitively expected[5], it holds under some non-trivial conditions given by the following theorem.

*Theorem 2:* Consider a full-rank multi-keyhole channel with the matrix $\mathbf{H}$ given by (3) under the following assumptions:

a) $\mathbf{h}_{tk}$ and $\mathbf{h}_{rk}$ are circular symmetric random vectors such that $E\|\mathbf{h}_{tk}\|^4 < \infty$ and $E\|\mathbf{h}_{rk}\|^4 < \infty$ $\forall k$,

b) the correlation matrix
$$\mathbf{C} = E\{vec(\mathbf{H})vec(\mathbf{H})^\dagger\} \\ = \sum_{k=1}^{M} |a_{k,M}|^2 \left(\mathbf{R}_{tk}^T \otimes \mathbf{R}_{rk}\right)$$

does not depend on $M$ and is non-singular[6], where $T$ denotes transposition, $\otimes$ denotes the Kronecker product, and $vec(\mathbf{H})$ creates a column vector by stacking the elements of $\mathbf{H}$ column-wise,

c) the following holds under normalization (4),
$$\lim_{M \to \infty} |a_{1,M}| = 0, \qquad (9)$$

where $|a_{1,M}| \geq |a_{2,M}| \geq ... \geq |a_{M,M}|$ are sorted in a non-increasing order.

Then,

(i) $\mathbf{H}$ is asymptotically circular symmetric complex Gaussian in distribution.

(ii) Let $\Delta_M = \sup_{\mathbf{x}} |F_M(\mathbf{x}) - \Phi(\mathbf{x})|$, where $F_M(\mathbf{x})$ is the cumulative distribution function (CDF) of $\mathbf{x} = vec(\mathbf{H})$ for given $M$ and $\Phi(\mathbf{x})$ is a Gaussian CDF with the with zero mean and the correlation matrix $\mathbf{C}$. Then, $\Delta_M \to 0$ as $M \to \infty$, i.e. $F_M(\mathbf{x})$ converges to $\Phi(\mathbf{x})$ uniformly, with at least the same rate as $|a_{1,M}| \to 0$.

*Proof:* see Appendix. ∎

Note that following Theorem 2, the multi-keyhole channel is asymptotically Rayleigh-fading, even though the sub-channels are not necessarily Rayleigh and/or uncorrelated.

---

[5] The multipath becomes richer with $M$ and so the channel distribution is closer to the Rayleigh one.

[6] Note that under normalization (4), $\mathbf{C}$ is the "average correlation matrix", averaged over $k = 1...M$. Thus, the assumption above is equivalent to considering a set of matrices $\mathbf{R}_{tk}^T \otimes \mathbf{R}_{rk}$, where the "average matrix" does not depend on the set size.

While (9) is a sufficient condition for Theorem 2 to hold, it provides only limited insight. Below we consider two equivalent conditions to obtain more insight.

*Corollary 2.1:* Condition (9) holds if and only if at least one of the conditions below is satisfied
$$\lim_{M \to \infty} \sum_{i=1}^{k} |a_{i,M}|^2 = 0, \quad \forall k < \infty, \qquad (10)$$
$$\lim_{M \to \infty} \sum_{i=1}^{M} |a_{i,M}|^{2+\delta} = 0, \quad \text{for some} \quad \delta > 0$$

*Proof:* see Appendix. ∎

In view of the normalization (4), Corollary 2.1 says that (9) holds if the power contribution of all keyholes is more or less the same, and that none of the keyholes contributes a significant part of the total power. Condition (9) does not hold when the number of non-zero keyholes is finite: $|a_{i,M}| > 0$, $i = 1...k$, and 0 otherwise, so that $\sum_{i=1}^{k} |a_{i,M}|^2 = 1 > 0$. Hence, a necessary condition for (9) to hold is that the number of non-zero keyholes increases to infinity with $M$.

The following corollary gives simple sufficient conditions for $\mathbf{C}$ to be non-singular.

*Corollary 2.2:* The correlation matrix
$$\mathbf{C} = \sum_{k=1}^{M} |a_{k,M}|^2 \left(\mathbf{R}_{tk}^T \otimes \mathbf{R}_{rk}\right)$$

is non-singular as $M \to \infty$ if either one of the following conditions is satisfied:

(i) all $\mathbf{R}_{tk}$, $\mathbf{R}_{rk}$ are full-rank, or

(ii) there is a set $S$ (either finite or infinite) of indices $k$ of singular matrices $\mathbf{R}_{tk}$, $\mathbf{R}_{rk}$, and
$$\lim_{M \to \infty} \sum_{k \notin S} |a_{k,M}|^2 > 0,$$

i.e. the power contribution of the keyholes with non-singular $\mathbf{R}_{tk}$, $\mathbf{R}_{rk}$ does not vanish as $M \to \infty$.

*Proof:* see Appendix. ∎

As an example, consider a multi-keyhole channel with $M$ identical keyholes, i.e. $|a_{k,M}| = \sqrt{1/M}$, $\mathbf{R}_t = \mathbf{R}_{tk}$ and $\mathbf{R}_r = \mathbf{R}_{rk}$ $\forall k$, and assume that they are non-singular. Clearly, (9) holds in this case, and under the normalization (4), $\mathbf{C} = \mathbf{R}_t^T \otimes \mathbf{R}_r$ is non-singular and does not depend on $M$. Therefore from Theorem 2(i), such a multi-keyhole channel converges in distribution to a Rayleigh-fading one as $M \to \infty$. Moreover, from Theorem 2(ii), the convergence is at least as $\sqrt{1/M}$.

Since the channel capacity is a continuous function of $\mathbf{H}$ (see (2)), the next corollary follows immediately from Theorem 2.

*Corollary 2.3:* Under the conditions of Theorem 2, the instantaneous capacity of an FRMK channel converges in distribution to that of the equivalent Rayleigh-fading channel.

*Proof:* by Slutsky Theorem [[42], Theorem 6a]. ∎

Fig. 3 compares the capacity distribution of the 2x2 multi-keyhole channel with $|a_{k,M}| = \sqrt{1/M}$ to that of the equivalent 2x2 Rayleigh-fading one. Correlation in both channels is simulated using the exponential model with correlation parameter $r = 0.5$ at both ends. Clearly, the outage capacity of the multi-keyhole channel increases with $M$ and approaches that of the equivalent Rayleigh-fading channel.

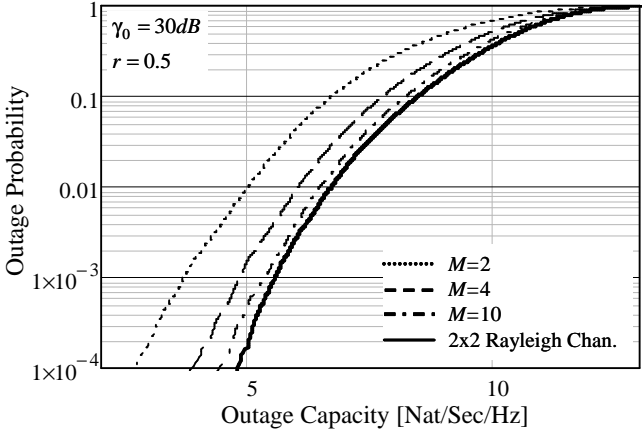

Fig. 3. Outage probability/capacity of 2x2 full-rank correlated Rayleigh-fading multi-keyhole channels.

Since the Tx and Rx ends are separated in the multi-keyhole channels (by the screen with keyholes, see Fig. 1), considering a Rayleigh-fading channel as a limiting case of the FRMK one provides a motivation for the popular Kronecker correlation model (see [29] for details on this model) as follows. Consider a multi-keyhole channel with $\mathbf{R}_t = \mathbf{R}_{tk}$ and $\mathbf{R}_r = \mathbf{R}_{rk}$, $\forall k$. It is straightforward to show, using (3) and (4), that

$$\begin{aligned} \mathbf{R}_r &= n_t^{-1} E\{\mathbf{H}\mathbf{H}^\dagger\} \\ \mathbf{R}_t &= n_r^{-1} E\{\mathbf{H}^\dagger\mathbf{H}\} \end{aligned} \qquad (11)$$

From (3), $\mathbf{H}$ can be represented in this case as

$$\mathbf{H} \propto \mathbf{R}_r^{1/2} \mathbf{G}_r \mathbf{A} \mathbf{G}_t^\dagger \mathbf{R}_t^{1/2}, \qquad (12)$$

where $\mathbf{G}_t$, $\mathbf{G}_r$ have i.i.d. Gaussian circular symmetric entries of unit variance. Since, under the conditions of Theorem 2, $\mathbf{G}_r \mathbf{A} \mathbf{G}_t^\dagger \to \mathbf{X}$ as $M \to \infty$, where $\mathbf{X}$ is an i.i.d. Gaussian circular symmetric matrix with unit-variance entries, the following holds

$$\mathbf{H} \xrightarrow{d} \mathbf{R}_r^{1/2} \mathbf{X} \mathbf{R}_t^{1/2}, \qquad (13)$$

where $\xrightarrow{d}$ denoted convergence in distribution as $M \to \infty$, and the right side of (13) is the Kronecker model for Rayleigh-fading channels. This clearly demonstrates that the Kronecker structure of the correlation is due to the separability of correlation-forming mechanisms into Tx and Rx parts.

The following theorem states that, under certain conditions, the capacity distribution of an FRMK channel is asymptotically Gaussian as $n_t, M \to \infty$. It is based on Theorem 2, and uses the fact that the capacity distribution of the Rayleigh-fading channel is asymptotically Gaussian [32].

*Theorem 3:* (i) Let $\mathbf{H}$ be the FRMK channel in (3), such that $\mathbf{h}_{tk}$ and $\mathbf{h}_{rk}$ are circular symmetric random vectors (not necessarily complex Gaussian), $n_t^{-2} E \|\mathbf{h}_{tk}\|^4 < \infty$, $n_r^{-2} E \|\mathbf{h}_{rk}\|^4 < \infty$, $\mathbf{R}_{tk} = \mathbf{R}_t$, $\mathbf{R}_{rk} = \mathbf{R}_r$, $\forall k$, and $\mathbf{R}_t$, $\mathbf{R}_r$ are positive definite and normalized, so that $n_t^{-1} tr\mathbf{R}_t = 1$ and $n_r^{-1} tr\mathbf{R}_r = 1$. If $n_t, M \to \infty$ at the rate such that

$$n_t^{3/2} \left\| \mathbf{R}_t^{-1/2} \right\|^3 \cdot |a_{1,M}| \to 0^7 \qquad (14)$$

$$\frac{\|\mathbf{R}_t\|_2}{\|\mathbf{R}_t\|} \to 0^8 \qquad (15)$$

where $\|\mathbf{R}_t\|_2$ is the spectral norm (largest eigenvalue) of $\mathbf{R}_t$.

Then the capacity distribution of such a multi-keyhole channel is asymptotically Gaussian with the following mean $\mu$ and variance $\sigma^2$:

$$\mu = \ln \det \left( \mathbf{I} + \frac{\gamma_0}{n_r} \mathbf{R}_r \right), \qquad (16)$$

$$\sigma^2 = \frac{1}{n_t^2} \|\mathbf{R}_t\|^2 \cdot \sum_{k=1}^{n_r} \left( \frac{\gamma_0 \lambda_k^r / n_r}{1 + \gamma_0 \lambda_k^r / n_r} \right)^2 \qquad (17)$$

At low per-eigenmode SNR, $\gamma_0 \lambda_k^r / n_r \ll 1$, where $\lambda_k^r$ are the eigenvalues of $\mathbf{R}_r$, $\mu$ and $\sigma^2$ are approximated by

$$\mu \approx \gamma_0, \qquad (18)$$

$$\sigma^2 \approx \gamma_0^2 \Psi_t \Psi_r, \qquad (19)$$

where $\Psi_t = n_t^{-2} \|\mathbf{R}_t\|^2$, $\Psi_r = n_r^{-2} \|\mathbf{R}_r\|^2$. The low SNR condition holds if $\gamma_0 \|\mathbf{R}_r\| / n_r \ll 1$, which is the case in "asymptotically uncorrelated" channels ($n_r^{-1} \|\mathbf{R}_r\| \to 0$).

(ii) Due to the symmetry in (5), this also hold true when Tx and Rx ends are exchanged.

*Proof:* see Appendix. ∎

Following the discussion in Section IV, the asymptotic mean capacity of the FRMK channel in (18) is independent of correlation and power imbalance (as measured by $\Psi_t, \Psi_r$). In contrast, the variance in (19) increases with it. However, under condition (15) of Theorem 3, $\Psi_t \to 0$ (i.e. the channel has to be "asymptotically uncorrelated" for the theorem to hold) and, therefore, $\sigma^2 \to 0$, so that the instantaneous capacity converges to the mean, $C \to \mu$, which is also know as "channel hardening" [30].

Following Theorem 3, the outage probability $P_{out} = \Pr\{C < R\}$, where the target rate $R = r\mu / \min(n_t, n_r, M)$ is expressed as a fraction of the mean capacity and $r$ is the multiplexing gain [41][56], can be compactly expressed as

$$\begin{aligned} P_{out} &= Q\left(\frac{\mu - R}{\sigma}\right) \\ &\approx Q\left(\frac{1 - r/\min(n_t, n_r, M)}{\sqrt{\Psi_t \Psi_r}}\right) \end{aligned} \qquad (20)$$

where $Q(x) = 1/\sqrt{2\pi} \cdot \int_x^\infty \exp\left(-t^2/2\right) dt$ is the Q-function, and the approximation holds at low SNR regime. We remark that, unlike [41], (20) gives an explicit closed-form relationship between $P_{out}$ and $r$ and also accounts for the effects of correlation and power imbalance in the channel. Note that $P_{out}$ does not depend on the SNR in the low SNR regime (since the rate $R$, the mean capacity $\mu$ and the standard deviation $\sigma$ are all proportional to the SNR) and increases with channel correlation and power imbalance in the region $P_{out} < 1/2$ (see Section IV for further discussion of correlation and power imbalance).

---

[7] This implies that $M$ has to increase to infinity much faster than $n_t$. For example, it is straightforward to show that when $\mathbf{R}_t = \mathbf{I}$, and $|a_{1,M}| = 1/\sqrt{M}$, (14) is equivalent to $n_t^6 / M \to 0$ as $n_t, M \to \infty$.

[8] This condition is elaborated in detail in [44].

## B. Rank-Deficient Multi-Keyhole Channel

Let us now consider a multi-keyhole channel where $M < min\{n_t, n_r\}$.

*Theorem 4:* Assume the following conditions hold

a) $\mathbf{h}_{tk} \propto \mathbf{R}_{tk}^{1/2} \mathbf{g}_{tk}$, $\mathbf{h}_{rk} \propto \mathbf{R}_{rk}^{1/2} \mathbf{g}_{rk}$, $k = 1...M$, where $\mathbf{g}_{tk}$ and $\mathbf{g}_{rk}$ are zero mean complex random vectors with independent entries (not necessarily Gaussian or identically distributed),

b) $m_{2+\delta}(i) < \infty$ and $m_2(i) > 0$ for all $i$ and some $\delta > 0$, where

$$m_\delta(i) = E\left(|g_i|^2 - E|g_i|^2\right)^\delta$$

is the central moment of $|g_i|^2$ of order $\delta$, and $g_i$ is the $i$-th entry of $\mathbf{g}_{tk}$, $\mathbf{g}_{rk}$,

c) $\lim_{n_t \to \infty} \frac{\|\mathbf{R}_{tk}\|_2}{\|\mathbf{R}_{tk}\|} = \lim_{n_t \to \infty} \frac{\|\mathbf{R}_{rk}\|_2}{\|\mathbf{R}_{rk}\|} = 0$, $\forall k$,

Then the instantaneous capacity (5) of the RDMK channel is asymptotically Gaussian as $n_t, n_r \to \infty$ with the following mean $\mu$ and the variance $\sigma^2$:

$$\mu = \sum_{k=1}^{M} \ln\left(1 + |a_{k,M}|^2 \gamma_0\right), \quad (21)$$

$$\sigma^2 = \sum_{k=1}^{M} \left(\frac{|a_{k,M}|^2 \gamma_0}{1 + |a_{k,M}|^2 \gamma_0}\right)^2 (\Psi_{tk} + \Psi_{rk}), \quad (22)$$

where $\Psi_{tk} = n_t^{-2} \|\mathbf{R}_{tk}\|^2$, $\Psi_{rk} = n_r^{-2} \|\mathbf{R}_{rk}\|^2$, and the $k$-th element of the sums in (21), (22) represents the contribution of $k$-th keyhole to the mean and variance of the instantaneous capacity.

*Proof:* using [[21], Theorems 4, 7], Comment 5 in [44], and Von-Neumann trace inequality [57]. ∎

Note that $\mu$ in (21) is not affected by the channel correlation, and $\sigma^2$ in (22) increases with $\Psi_{tk}$, $\Psi_{rk}$. Under condition (c), $\Psi_{tk}, \Psi_{rk} \to 0$, i.e. $\sigma^2 \to 0$, so that similarly to the FRMK channel, the instantaneous capacity converges to the mean, $C \to \mu$.

A number of approximations of (21) and (22) are in order:
*Low SNR regime* ($|a_{k,M}|^2 \gamma_0 \ll 1$, $\forall k$):

$$\mu \approx \gamma_0 \quad (23)$$
$$\sigma^2 \approx \gamma_0^2 \sum_{k=1}^{M} |a_{k,M}|^2 (\Psi_{tk} + \Psi_{rk})$$

*High SNR regime* ($|a_{k,M}|^2 \gamma_0 \gg 1$, $\forall k$):

$$\mu \approx M \ln \gamma_0 + \sum_{k=1}^{M} \ln |a_{k,M}|^2$$
$$\sigma^2 \approx \sum_{k=1}^{M} (\Psi_{tk} + \Psi_{rk}) \quad (24)$$

Using these approximations, the outage probability of the RDMK channel in the low SNR regime is

$$P_{out} \approx Q\left(\frac{1 - r/M}{\sqrt{\sum_{k=1}^{M} |a_{k,M}|^2 (\Psi_{tk} + \Psi_{rk})}}\right), \quad (25)$$

i.e. it is also independent of the SNR and increases with channel correlation and power imbalance in the region $P_{out} < 1/2$.

The fact that the asymptotic outage probability and also capacity of FRMK and RDMK channels are the functions of $\|\mathbf{R}_t\|$ and $\|\mathbf{R}_r\|$ (see (16)-(25)) motivates the following proposition:

*Proposition 1:* Asymptotically, the channel correlation affects the outage capacity through the Frobenius norm of the correlation matrices, i.e. even though some $\mathbf{R}_1$ and $\mathbf{R}_2$ (at either end) are different, they affect the capacity in the same way if $\|\mathbf{R}_1\| = \|\mathbf{R}_2\|$.

Proposition 1 suggests a simple and well-tractable measure of correlation, whose properties are studied in the next section.

## IV. SCALAR MEASURES OF CORRELATION AND POWER IMBALANCE

Consider a correlation matrix $\mathbf{R}$ at either Tx or Rx end. Let $\mathbf{R} \in \mathcal{M}$, a set of all $n \times n$ normalized correlation matrices, $tr(\mathbf{R}) = n$. It is straightforward to show that $n^{-1} \|\mathbf{R}\|$ is bounded,

$$\frac{1}{\sqrt{n}} \leq \frac{1}{n} \|\mathbf{R}\| \leq 1, \quad (26)$$

where the lower bound is achieved when the channel is uncorrelated with the same power at each Tx(Rx) antenna, i.e. $\mathbf{R} = \mathbf{I}$, and the upper bound is achieved when the channel is fully correlated, i.e. $\mathbf{R}$ has a single non-zero eigenvalue. There are two major effects that can increase $n^{-1} \|\mathbf{R}\|$: (i) non-uniform power distribution across the antennas (also termed power imbalance), and (ii) non-zero correlation. To analyze these effects separately, we split $\mathbf{R} \in \mathcal{M}$ into a sum of two matrices as follows:

$$\mathbf{R} = \mathbf{K} + \mathbf{P}, \quad (27)$$

where $\mathbf{P} = diag\{\mathbf{R}\} - \mathbf{I}$ and $\mathbf{K} = \mathbf{R} - \mathbf{P}$; $diag\{\mathbf{R}\}$ is the diagonal matrix whose main diagonal is that of $\mathbf{R}$. $\mathbf{P}$ and $\mathbf{K}$ account for the power imbalance and the correlation respectively. Since for any $\mathbf{R} \in \mathcal{M}$, $tr(\mathbf{K}) = n$ and $tr(\mathbf{P}) = 0$, it is straightforward to show that the decomposition (27) is norm-orthogonal, i.e.

$$\|\mathbf{R}\|^2 = \|\mathbf{K}\|^2 + \|\mathbf{P}\|^2, \quad (28)$$

and $n^{-1} \|\mathbf{P}\|$ is bounded by

$$0 \leq \frac{1}{n} \|\mathbf{P}\| \leq \sqrt{1 - \frac{1}{n}}, \quad (29)$$

where the lower bound is achieved when all antennas have the same power (no power imbalance), i.e. $\mathbf{P}$ is a zero matrix, or equivalently $diag\{\mathbf{R}\} = \mathbf{I}$, and the upper bound is achieved when there is only one active antenna, i.e. there is only one

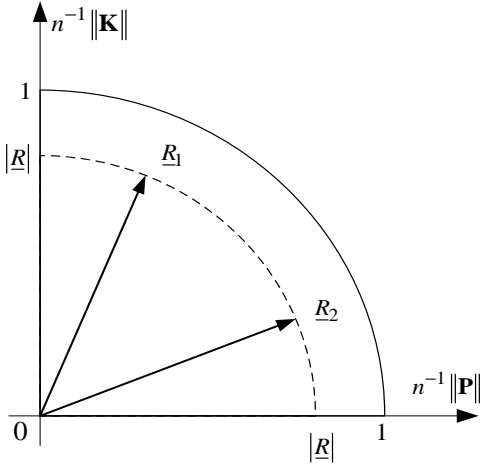

Fig. 4.  Geometrical interpretation of power imbalance and correlation effects.

non-zero diagonal entry in $\mathbf{R}$. Using (26) and (28), it is straightforward to show that

$$\frac{1}{\sqrt{n}} \leq \frac{1}{n} \|\mathbf{K}\| \leq 1, \tag{30}$$

where the lower bound is achieved when the channel is uncorrelated, $\mathbf{K} = \mathbf{I}$, and the upper bound is achieved when the channel is fully correlated. The following definitions are motivated by the discussion above and Proposition 1.

*Definition 1:* A MIMO channel with correlation matrix $\mathbf{R}_1 \in \mathcal{M}$ is said to be equally or more correlated than that with $\mathbf{R}_2 \in \mathcal{M}$, if

$$\|\mathbf{K}_1\| \geq \|\mathbf{K}_2\|, \tag{31}$$

where $\mathbf{K}_1$ and $\mathbf{K}_2$ correspond to $\mathbf{R}_1$ and $\mathbf{R}_2$ via (27).

*Definition 2:* A MIMO channel with correlation matrix $\mathbf{R}_1 \in \mathcal{M}$ has higher power imbalance than that with $\mathbf{R}_2 \in \mathcal{M}$ if

$$\|\mathbf{P}_1\| \geq \|\mathbf{P}_2\|, \tag{32}$$

where $\mathbf{P}_1$ and $\mathbf{P}_2$ correspond to $\mathbf{R}_1$ and $\mathbf{R}_2$ via (27).

From (29) and (30), the measures of correlation and power imbalance are bounded as $n^{-1}\|\mathbf{K}\| \in (0;1]$, $n^{-1}\|\mathbf{P}\| \in [0;1)$ when $n \to \infty$. Note also that due to the properties of the Frobenius norm [43], the measure is invariant under unitary transformation of $\mathbf{R}$. Since the eigenvalue decomposition $\mathbf{R} = \mathbf{U}\mathbf{\Lambda}\mathbf{U}^\dagger$ is a particular case of a unitary transformation, the impact of correlation on the asymptotic channel capacity is the same whether the correlation matrix is $\mathbf{R}$ or $\mathbf{\Lambda}$. Since the latter also describes channels with no correlation and non-uniform power distribution, the effects of correlation and power imbalance are indistinguishable in the eigenspace of correlation matrices. On the contrary, the decomposition in (27) clearly separates these effects.

To get some insight, consider a simple geometrical interpretation of Definitions 1 and 2 shown in Fig. 4. From (28), $n^{-1}\|\mathbf{K}\|$ and $n^{-1}\|\mathbf{P}\|$ create an orthonormal basis in a vector space, and the measure of correlation and power imbalance is a mapping of $\mathcal{M}$ onto a circle sector in that basis. The channel correlation matrix $\mathbf{R}$ is represented by a two dimensional vector $\mathbf{r}$ (see Fig. 4) such that

$$|\mathbf{r}| = \frac{1}{n}\|\mathbf{R}\|; \ angle\{R\} = \tan^{-1}\left(\|\mathbf{P}\|/\|\mathbf{K}\|\right) \tag{33}$$

Following Proposition 1, the asymptotic outage capacity is affected by the length of $\mathbf{r}$, but not by its angle. Consider two channels with correlation matrices represented by the vectors $R_1$ and $R_2$, such that $|R_1| = |R_2| = |R|$ (see Fig. 4). Following Definitions 1 and 2, the channel with $R_1$ is more correlated than that with $R_2$. In turn, the channel with $R_2$ has more power imbalance across antennas. Nonetheless, the asymptotic outage capacity of both channels is the same, i.e. the power imbalance and correlation between antennas have the same impact on the asymptotic capacity distribution of MIMO channels, if $|R_1| = |R_2|$.

We note that the measure $n^{-1}\|\mathbf{R}\|$ also characterizes the impact of correlation on the mean capacity and diversity gain in Rayleigh-fading channels, as shown in [34][35][45]. The results above show that it also applies to the outage capacity of a broad class of multi-keyhole channels, and also characterizes the effect of power imbalance. We thus conclude that this measure has a high degree of universality in the characterization of channel performance.

Unlike the measures based on majorization theory [37], $n^{-1}\|\mathbf{R}\|$ has full ordering property (any two channels can be compared, without exceptions). Moreover, there is a direct relationship between this measure and that in [37] as indicated by the following theorem.

*Theorem 5:* Let $\mathcal{M}_M$ be a subset in $\mathcal{M}$ of all correlation matrices which can be majorized[9]. Then, for any $\mathbf{R}_1$, $\mathbf{R}_2 \in \mathcal{M}_M$, $\mathbf{R}_1 \succ \mathbf{R}_2$ if and only if $n^{-1}\|\mathbf{R}_1\| \geq n^{-1}\|\mathbf{R}_2\|$.[10]

*Proof:* see Appendix. ∎

Consider, as an example, the exponential correlation matrix [40], which has been successfully used to model correlation in several problems [40][46]. In this model, the elements of $\mathbf{R}$ are given by

$$R_{km} = \begin{cases} r^{m-k}; & m \geq k \\ \bar{r}^{k-m}; & m < k \end{cases}, \ |r| < 1, \tag{34}$$

where $r$ is a correlation parameter (the correlation between two adjacent antennas), and $\bar{r}$ is the complex conjugate of $r$. From (27), $\|\mathbf{K}\| = \|\mathbf{R}\|$ and $\|\mathbf{P}\| = 0$, i.e. this model does not capture the effect of power imbalance, but the correlation only. From [21],

$$\frac{1}{n^2}\|\mathbf{R}\|^2 = \frac{1}{n} \cdot \frac{1+|r|^2}{1-|r|^2} + o(n^{-1}), \quad |r| < 1, \tag{35}$$

where $o(x)$ is the value such that $\lim_{x \to 0} o(x)/x = 0$. Hence for large $n$, the measure of correlation in this case increases monotonically with $|r|$, which supports Definition 1. Moreover, $n^{-2}\|\mathbf{R}\|^2$ monotonically decreases with $n$ and eventually converges to zero as $n \to \infty$, i.e. even though the correlation between adjacent antennas may be high, an increase in the number of antennas reduces the measure of correlation due to smaller correlation between distant antennas. Note that this property of the channel to be "asymptotically uncorrelated" is a condition for Theorems 1, 3 and 4 to hold.

---

[9] A correlation matrix $\mathbf{R}_1$ is said to majorize (more correlated than) $\mathbf{R}_2$ and denoted by $\mathbf{R}_1 \succ \mathbf{R}_2$, if $\sum_{k=1}^{m} \lambda_k^{(1)} \geq \sum_{k=1}^{m} \lambda_k^{(2)}$ for all $m = 1...n$, where $\lambda_k^{(1)}$ and $\lambda_k^{(2)}$ are the eigenvalues of $\mathbf{R}_1$ and $\mathbf{R}_2$ respectively sorted in the descending order [37].

[10] The same result was independently obtained in [45].

Another model with similar asymptotic behavior is the quadratic exponential correlation model. This is a physically based model which represents the scenario with a Gaussian profile of multipath angle-of-arrival [47]. As in the exponential correlation model, the measure of correlation and power imbalance in this case increases monotonically with $|r|$, which supports Definition 1, and monotonically converges to zero as $n \to \infty$ (see [21] for more details). It is straightforward to show that both exponential or quadratic exponential correlation models satisfy the relevant conditions of Theorems 3, 4. However, the latter are not satisfied when $\mathbf{R}$ is given by the uniform correlation model [48], where the correlation between any pair of antennas is the same.

## V. IMPACT OF CORRELATION AND POWER IMBALANCE ON THE OUTAGE CAPACITY

Using Definitions 1 and 2 above, we note that the mean capacity of RDMK channels and FRMK ones in the low SNR regime (see (21) and (18)) is independent of the correlation and power imbalance. The variance, in turn, increases with it (see (22) and (19)). The conditions for the asymptotic instantaneous capacity to be Gaussian essentially require the channel to be "asymptotically uncorrelated", i.e. $\Psi_{t(r)} \to 0$.

Using (20), the outage capacity $C_\varepsilon$ can be expressed as,

$$C_\varepsilon = \mu - \sigma Q^{-1}(\varepsilon), \qquad (36)$$

where $Q^{-1}$ is the inverse of the Q-function. Note that $2^{nd}$ term is positive, i.e. $C_\varepsilon < \mu$, if $\varepsilon < 1/2$, and negative, i.e. $C_\varepsilon > \mu$, otherwise. Only the $1^{st}$ case is of practical importance, which is considered in the following proposition.

*Proposition 2:* Under the conditions of Theorems 3 and 4 the outage capacities of FRMK and RDMK channels at low SNR regime are

$$C_\varepsilon \approx \gamma_0 \left( 1 - \sqrt{\Psi_t \Psi_r} Q^{-1}(\varepsilon) \right) \qquad (37)$$

$$C_\varepsilon \approx \gamma_0 \left( 1 - Q^{-1}(\varepsilon) \sqrt{\sum_{k=1}^{M} |a_{k,M}|^2 \left( \Psi_{tk} + \Psi_{rk} \right)} \right) \qquad (38)$$

respectively.

*Proof:* (37) is obtained using (18) and (19). (38) follows from (23). ∎

Clearly, the outage capacity decreases, in both cases, with the measure of correlation and power imbalance at both ends. Note that the impacts of the target outage probability $\varepsilon$, the SNR and the correlation/power imbalance are clearly separated in (37), (38), e.g. the outage capacity is proportional to the average SNR and the capacity loss is proportional to $Q^{-1}(\varepsilon)$.

While the above analysis is based on the asymptotic assumption $n_t, n_r \to \infty$, numerical simulations show that Theorems 3 and 4 adequately characterize the impact of correlation on the outage capacity of multi-keyhole channels with a moderate number of antennas as well.

## VI. APPLICATIONS

In this section, we address some problems whose solution for a finite number of antennas is associated with significant mathematical complexity. We show that in the asymptotic regime, these problems are well-tractable, and obtain compact closed-form solutions.

### A. Telatar's Conjecture

*Conjecture 1 (Telatar [1]):* Consider the outage probability of a MIMO block fading channel with full CSI at Rx end but no CSI at the Tx end,

$$P_{out} = \min_{\substack{tr\mathbf{R}_x \leq m \\ \mathbf{R}_x \geq \mathbf{0}}} \Pr \left\{ \ln \det \left( \mathbf{I} + \frac{\gamma_0}{n_t n_r} \mathbf{H} \mathbf{R}_x \mathbf{H}^\dagger \right) \leq R \right\} \qquad (39)$$

where $\mathbf{R}_x = \frac{k}{P_T} E\{\mathbf{x}\mathbf{x}^\dagger\} \geq 0$ is the normalized Tx covariance matrix, and $\mathbf{A} \geq \mathbf{0}$ denotes positive semidefinite matrix $\mathbf{A}$. For an i.i.d. Rayleigh-fading channel the outage probability is achieved when

$$\mathbf{R}_x = \mathbf{Q}^* = \begin{bmatrix} \mathbf{I}_{k \times k} & \mathbf{0} \\ \mathbf{0} & \mathbf{0} \end{bmatrix}, \qquad (40)$$

where the number of active antennas $k$ depends on the rate: higher the rate (i.e., higher the outage probability), smaller the $k$. ∎

Telatar's conjecture has been proven for multiple-input-single-output MISO Rayleigh-fading channels in [36][49]. The theorem below affirms the conjecture for semi-correlated multi-keyhole channels with large number of antennas.

*Theorem 6:* Consider a semi-correlated multi-keyhole channel, uncorrelated at the Tx end ($\mathbf{R}_t = \mathbf{I}$). Under the conditions of Theorems 3 and 4, the optimal Tx covariance matrix $\mathbf{R}_x = \mathbf{Q}^*$, which minimizes the outage probability at the region $P_{out} < 1/2$, is as in (40) with $k = n_t$.

*Proof:* (i) Let $\mathbf{Q}$ be a positive semi-definite matrix. It follows from (1)-(3) that the outage capacity of the multi-keyhole channel with $\mathbf{R}_t = \mathbf{I}$, $\mathbf{R}_x = \mathbf{Q}$ equals to that with $\mathbf{R}_t = \mathbf{Q}$, $\mathbf{R}_x = \mathbf{I}$, i.e. $\mathbf{R}_t$ can be treated as either the channel Tx correlation matrix or the Tx signal covariance matrix.

(ii) As $n_t, n_r \to \infty$, the outage probability in the Tx-correlated multi-keyhole channel increases with $n_t^{-1} \|\mathbf{R}_t\|$ at the region $P_{out} < 1/2$ (see Theorems 3 and 4 for FRMK and RDMK channels respectively). Combining (i) and (ii), we conclude that the outage probability in the semi-correlated multi-keyhole channel ($\mathbf{R}_t = \mathbf{I}$) increases with $n_t^{-1} \|\mathbf{R}_x\|$, so that the outage probability achieves the minimum when $n_t^{-1} \|\mathbf{R}_x\|$ is minimal. From (26), this is achieved when $\mathbf{R}_x = \mathbf{I}$. ∎

Following the same argument, $\mathbf{R}_x = \mathbf{I}$ also maximizes the mean capacity and the diversity order of semi-correlated Rayleigh-fading channels (uncorrelated at the Tx end), since both characteristics decrease with $n_t^{-1} \|\mathbf{R}_t\|$ [34][35][45]. The following corollary shows that this also holds for the outage capacity of a Rayleigh-fading channel.

*Corollary 6.1:* Consider a semi-correlated Rayleigh-fading channel (uncorrelated at the Tx end, $\mathbf{R}_t = \mathbf{I}$). Under the conditions of Theorem 3, the optimal Tx covariance matrix, which minimizes the outage probability and, thus, maximizes the outage capacity, at the region $P_{out} < 0.5$, is as in (40) with $k = n_t$.

*Proof:* by Theorem 6 and using the fact that the Rayleigh-fading channel is a special case of the multi-keyhole one when $M \to \infty$ (see Theorem 2). ■

Corollary 6.1 is in fact a generalization of Telatar's conjecture for the semi-correlated Rayleigh channels (recall that the original conjecture applies to i.i.d. channels only). The intuition behind it is that the multi-keyhole channel and thus the corresponding Rayleigh channel are separated into independent Tx and Rx parts so that correlation at the Rx end cannot affect the optimal covariance matrix at the Tx end.

### B. Throughput Gain and Feedback Rate in Multi-User Channels

Let us consider a communications environment with multiple users and a single base station. Assume that there is no direct link between each user and the base station, but only via $M$ parallel "amplify and forward" relay nodes (i.e. via $M$ dual-hop paths, as in Fig. 1). In many practically-important cases the relay noise can be neglected (see [61] for details), and therefore, the relay channel in these cases is well approximated by the multi-keyhole model in (3).

There is a number of transmission scheduling algorithms that allow increased data throughput in multi-user channels. An efficient method to estimate throughput gain due to a scheduling and the corresponding feedback rate has been proposed in [30], assuming that instantaneous capacity of a user-base station link is a Gaussian random variable. Following Theorems 3 and 4, the method in [30] applies also to multi-user multi-keyhole/relay channels, when the number of antennas is large enough to apply the Gaussian approximation of the instantaneous capacity with reasonable accuracy. In particular, the throughput gain and the feedback rate are obtained in a straightforward way by substituting the asymptotic moments in (16), (17), (21) and (22) in [[30], eq. 33 and 49] respectively (see [51] for more details).

## VII. CONCLUSION

A profound reason to study multi-keyhole channels is not only due to the fact that they model a number of practically-important propagation scenarios, including relay channels in the amplify-and-forward mode, but also because they are of considerable interest from the information-theoretic point of view as a transition model that spans a wide spectrum of MIMO channels, from the rank-one single-keyhole to full-rank Rayleigh-fading. Investigation of multi-keyhole channels provides an insight into rank-deficient and full-rank (not necessarily Rayleigh-fading) channels, and gives new insight into correlation-forming mechanisms. In particular, considering a Rayleigh-fading channel as a multi-keyhole one with a large number of keyholes provides an additional motivation for the popular Kronecker model. The outage capacity analysis of the multi-keyhole channels with a large number of antennas shows that the spatial correlation as well as antenna power imbalance can dramatically increase the outage probability (which cannot be captured using the popular diversity-multiplexing framework). Asymptotically (in number of antennas), the impact of correlation and power imbalance on the outage capacity is characterized by the Frobenius norm of the correlation matrices at both ends, which motivates a simple and well tractable scalar measure of correlation and power imbalance in MIMO channels. Finally, the asymptotic analysis shows that the Gaussian approximation of the capacity distribution has a high degree of universality and applies to a wide class of MIMO channels, far beyond the canonic Rayleigh-fading one. This asymptotic property allows one to obtain compact solutions for a number of problems, including those in the multi-user communications, for a broad class of fading channels.

## ACKNOWLEDGEMENT

The authors would like to thank the late Professor Vidmantas Bentkus for valuable comments on Central Limit Theorems, and the editor and the reviewers for many constructive suggestions.

## APPENDIX

### A. Proof of Theorem 1:

We start with the following lemmas:

*Lemma 1:* Let $\beta = \|\mathbf{h}\|^2$ be a generalized $\chi^2$ random variable characterized by correlation matrix $\mathbf{R} = E\{\mathbf{h}\mathbf{h}^\dagger\}$ [50]. If $\lim_{n \to \infty} n^{-1} tr\mathbf{R}$ and $\lim_{n \to \infty} n^{-2} \|\mathbf{R}\|^2$ are finite, then as $n \to \infty$, $n^{-1}\beta$ is Gaussian in distribution with the mean $\mu = n^{-1} tr\mathbf{R}$ and variance $\sigma^2 = n^{-2} \|\mathbf{R}\|^2$.

*Proof:* see the proof of Lemma B in [[21], Appendix B]. ■

*Lemma 2:* Let $\mathbf{H}$ be an $n \times M$ random matrix with $M$ mutually independent columns $\mathbf{h}_1 ... \mathbf{h}_M$, such that $\mathbf{h}_k, k = 1 ... M$, is a Gaussian circularly symmetric vector with correlation matrix $\mathbf{R}_k = E\{\mathbf{h}_k \mathbf{h}_k^\dagger\}$. If $n^{-1} tr\mathbf{R}_k = 1$ and $n^{-1} \|\mathbf{R}_k\| \to 0$ as $n \to \infty$, then $\mathbf{H}^\dagger \mathbf{H}/n \overset{q.m.}{\to} \mathbf{I}$ and therefore $\mathbf{H}^\dagger \mathbf{H}/n \overset{p}{\to} \mathbf{I}$ as $n \to \infty$, where $\mathbf{I}$ is an $M \times M$ identity matrix, $\overset{q.m.}{\to}$ denotes convergence in quadratic mean, i.e.

$$\lim_{n \to \infty} E \left\| \mathbf{H}^\dagger \mathbf{H}/n - \mathbf{I} \right\|^2 = 0,$$

and $\overset{p}{\to}$ denotes convergence in probability, i.e.

$$\lim_{n \to \infty} \Pr \left\{ \left\| \mathbf{H}^\dagger \mathbf{H}/n - \mathbf{I} \right\| \geq \varepsilon \right\} = 0, \quad \forall \varepsilon > 0.$$

*Proof:* Since $\mathbf{h}_1 ... \mathbf{h}_M$ are mutually independent, assume, without loss of generality, that $\mathbf{H}$ is normalized such that $E\{\mathbf{H}^\dagger \mathbf{H}/n\} = \mathbf{I}$. From Chebyshev inequality, for any $\varepsilon > 0$

$$\Pr \left\{ \left\| \frac{1}{n} \mathbf{H}^\dagger \mathbf{H} - \mathbf{I} \right\| \geq \varepsilon \right\} \leq \varepsilon^{-2} \cdot E \left\| \frac{1}{n} \mathbf{H}^\dagger \mathbf{H} - \mathbf{I} \right\|^2, \quad (41)$$

where

$$E\left\|\frac{1}{n}\mathbf{H}^\dagger\mathbf{H} - \mathbf{I}\right\|^2 = \frac{1}{n^2} tr E\{\mathbf{H}\mathbf{H}^\dagger\mathbf{H}\mathbf{H}^\dagger\} - M$$
$$= \frac{1}{n^2} \sum_{m=1}^{M} \sum_{k=1}^{M} tr E\{\mathbf{h}_m\mathbf{h}_m^\dagger\mathbf{h}_k\mathbf{h}_k^\dagger\} - M \quad (42)$$

Since $\sqrt{tr\{\mathbf{h}_k\mathbf{h}_k^\dagger\mathbf{h}_k\mathbf{h}_k^\dagger\}} = \|\mathbf{h}_k\|^2$ is a generalized $\chi^2$ random variable characterized by $\mathbf{R}_k$. Thus, using Lemma 1:

$$\frac{1}{n^2} tr E\{\mathbf{h}_m\mathbf{h}_m^\dagger\mathbf{h}_k\mathbf{h}_k^\dagger\} = \begin{cases} n^{-2}\|\mathbf{R}_k\|^2 + 1; & k = m \\ n^{-2} tr[\mathbf{R}_m\mathbf{R}_k]; & k \neq m \end{cases} \quad (43)$$

Substituting (43) in (42), one obtains

$$E\left\|\frac{1}{n}\mathbf{H}^\dagger\mathbf{H} - \mathbf{I}\right\|^2 = n^{-2}\sum_{m=1}^{M}\sum_{k=1}^{M} tr[\mathbf{R}_m\mathbf{R}_k] \quad (44)$$

Using Cauchy-Schwartz's inequality

$$|tr[\mathbf{R}_m\mathbf{R}_k]| \leq \|\mathbf{R}_m\| \cdot \|\mathbf{R}_k\|.$$

Therefore, if $n^{-1}\|\mathbf{R}_k\| \to 0$, then $n^{-2}tr[\mathbf{R}_m\mathbf{R}_k] \to 0$ as $n \to \infty$. Using (44)

$$E\left\|\frac{1}{n}\mathbf{H}^\dagger\mathbf{H} - \mathbf{I}\right\|^2 \to 0, \quad (45)$$

or equivalently using (41) for any $\varepsilon > 0$

$$\Pr\left\{\left\|\frac{1}{n}\mathbf{H}^\dagger\mathbf{H} - \mathbf{I}\right\| \geq \varepsilon\right\} \to 0 \quad (46)$$

∎

(i) Under the conditions of Lemma 2, for $\mathbf{H} = \mathbf{H}_t$ and $n = n_t$

$$\mathbf{B}_t = \frac{1}{n_t}\mathbf{H}_t^\dagger\mathbf{H}_t \xrightarrow{p} \mathbf{I} \quad \text{as} \quad n_t \to \infty \quad (47)$$

Since (5) is a continuous function of $\mathbf{B}_t$, from Slutsky Theorem [[42], Theorem 6'(a)], as $n_t \to \infty$

$$\begin{aligned} C &= \ln\det\left(\mathbf{I} + \gamma_0 \mathbf{B}_r \mathbf{A} \mathbf{B}_t \mathbf{A}^\dagger\right) \\ &\xrightarrow{p} \ln\det\left(\mathbf{I} + \gamma_0 \mathbf{B}_r \mathbf{A}\mathbf{A}^\dagger\right) \\ &= \ln\det\left(\mathbf{I} + \frac{\gamma_0}{n_r}\mathbf{H}_r \mathbf{A}\mathbf{A}^\dagger\mathbf{H}_r^\dagger\right), \end{aligned} \quad (48)$$

where the right side is the instantaneous capacity of an $M \times n_r$ equivalent Rayleigh-fading channel with channel matrix $\mathbf{H}_r$ and power allocation matrix $\mathbf{A}\mathbf{A}^\dagger$.

(ii) By the same argument as for (i), as $n_r \to \infty$

$$\mathbf{B}_r = \frac{1}{n_r}\mathbf{H}_r^\dagger\mathbf{H}_r \xrightarrow{p} \mathbf{I} \quad (49)$$

From Slutsky Theorem [[42], Theorem 6'(a)], as $n_r \to \infty$

$$\begin{aligned} C &= \ln\det\left(\mathbf{I} + \gamma_0 \mathbf{B}_r \mathbf{A} \mathbf{B}_t \mathbf{A}^\dagger\right) \\ &\xrightarrow{p} \ln\det\left(\mathbf{I} + \gamma_0 \mathbf{A}^\dagger \mathbf{A} \mathbf{B}_t\right) \\ &= \ln\det\left(\mathbf{I} + \frac{\gamma_0}{n_r}\mathbf{H}_t \mathbf{A}^\dagger \mathbf{A} \mathbf{H}_t^\dagger\right), \end{aligned} \quad (50)$$

where the right side is the instantaneous capacity of an $n_t \times M$ equivalent Rayleigh-fading channel with the channel matrix $\mathbf{H}_t$, and power allocation matrix $\mathbf{A}^\dagger\mathbf{A}$. The equality of outage probabilities follows from the convergence in probability [42]. ∎

### B. Proof of Theorem 2:

We start with the following theorem and corollary.

*Theorem 7 (Bentkus [52]):* Let $\mathbf{s} = \sum_{k=1}^{n} \mathbf{x}_k$, where $\mathbf{x}_1...\mathbf{x}_n$ are mutually independent random vectors taking values in $\mathcal{R}^d$ such that $E\{\mathbf{x}_k\} = 0$, $\forall k$, and $\mathbf{C} = E\{\mathbf{s} \cdot \mathbf{s}^T\}$ is invertible. Then, as $n \to \infty$, $\mathbf{s}$ is asymptotically Gaussian in distribution with zero mean and covariance matrix $\mathbf{C}$ if

$$\lim_{n\to\infty} \sum_{k=1}^{n} E\left\|\mathbf{C}^{-1/2}\mathbf{x}_k\right\|^3 = 0 \quad (51)$$

Moreover, let $\Delta_n = \sup_{\mathbf{x}} |F_n(\mathbf{x}) - \Phi(\mathbf{x})|$, where $F_n(\mathbf{x})$ is the CDF of $\mathbf{s}$ and $\Phi(\mathbf{x})$ is a Gaussian CDF with the same mean and variance as of $\mathbf{s}$, then $\Delta_n \to 0$ with the same rate as

$$\sum_{k=1}^{n} E\left\|\mathbf{C}^{-1/2}\mathbf{x}_k\right\|^3 \to 0.$$

A generalization of Theorem 7 for a complex case is given by the following corollary.

*Corollary 7.1:* Let $\mathbf{s} = \sum_{k=1}^{n} \mathbf{x}_k$, where $\mathbf{x}_1...\mathbf{x}_n$ are mutually independent circularly symmetric random vectors taking values in $\mathcal{C}^d$ such that $E\{\mathbf{x}_k\} = 0$, $\forall k$, and $\mathbf{C} = E\{\mathbf{s} \cdot \mathbf{s}^\dagger\}$ is invertible. Then Theorem 7 holds.

*Proof:* A proof is standard, based on $\mathcal{C}^d \to \mathcal{R}^{2d}$ mapping, and follows immediately from the properties of circular symmetric random vectors, see [[1], Lemma 1]. ∎

(i) Let $\mathbf{H}$ be a matrix of a multi-keyhole channel defined in (3). It is straightforward to show that

$$vec(\mathbf{H}) = \sum_{k=1}^{M} \mathbf{x}_k, \quad (52)$$

where $\mathbf{x}_k = a_{k,M} vec(\mathbf{h}_{rk}\mathbf{h}_{tk}^\dagger)$. Since $\mathbf{h}_{tk}$ and $\mathbf{h}_{rk}$ are mutually independent, $\mathbf{x}_k$ are mutually independent circular symmetric random vectors. Thus, following Theorem 7 and Corollary 7.1, $vec(\mathbf{H})$ is asymptotically circular symmetric Gaussian as $M \to \infty$ if

$$\lim_{M\to\infty} \sum_{k=1}^{M} E\left\|\mathbf{C}^{-1/2}\mathbf{x}_k\right\|^3 = 0, \quad (53)$$

where

$$\mathbf{C} = E\left\{vec(\mathbf{H}) \cdot vec(\mathbf{H})^\dagger\right\} = \sum_{k=1}^{M} |a_{k,M}|^2 \mathbf{R}_{tk}^T \otimes \mathbf{R}_{rk} \quad (54)$$

Consider the following upper bounds

$$\begin{aligned} \sum_{k=1}^{M} E\left\|\mathbf{C}^{-1/2}\mathbf{x}_k\right\|^3 &\leq \sum_{k=1}^{M} \left(E\left\|\mathbf{C}^{-1/2}\mathbf{x}_k\right\|^4\right)^{3/4} \\ &\leq \left\|\mathbf{C}^{-1/2}\right\|^3 \cdot \sum_{k=1}^{M} \left(E\|\mathbf{x}_k\|^4\right)^{3/4} \\ &= \left\|\mathbf{C}^{-1/2}\right\|^3 \cdot \sum_{k=1}^{M} |a_{k,M}|^3 \left(E\|\mathbf{h}_{tk}\|^4 \cdot E\|\mathbf{h}_{rk}\|^4\right)^{3/4} \\ &\leq \left\|\mathbf{C}^{-1/2}\right\|^3 \cdot \max_k \left(E\|\mathbf{h}_{tk}\|^4 \cdot E\|\mathbf{h}_{rk}\|^4\right)^{3/4} \sum_{k=1}^{M} |a_{k,M}|^3 \\ &\leq \left\|\mathbf{C}^{-1/2}\right\|^3 \cdot \max_k \left(E\|\mathbf{h}_{tk}\|^4 \cdot E\|\mathbf{h}_{rk}\|^4\right)^{3/4} \cdot |a_{1,M}| \end{aligned} \quad (55)$$

where the inequalities are due to Lyapunov [[53], Theorem 3.4.1], Cauchy-Schwartz inequalities and (57), and the equality

follows from $\mathbf{x}_k = a_{k,M} vec(\mathbf{h}_{rk}\mathbf{h}_{tk}^\dagger)$. Under the adopted assumptions, $\|\mathbf{C}^{-1/2}\|$ and

$$\max_k \left( E \|\mathbf{h}_{tk}\|^4 \cdot E \|\mathbf{h}_{rk}\|^4 \right)^{3/4}$$

are finite. Thus, (53) holds if

$$\lim_{M \to \infty} |a_{1,M}| = 0, \quad (56)$$

From Theorem 7 and Corollary 7.1, $vec(\mathbf{H})$ is asymptotically circular symmetric Gaussian as $M \to \infty$.

(ii) Let $\Delta_M = \sup_\mathbf{x} |F_M(\mathbf{x}) - \Phi(\mathbf{x})|$, where $F_M(\mathbf{x})$ is the CDF of $vec(\mathbf{H})$ and $\Phi(\mathbf{x})$ is a Gaussian CDF with the same mean and variance as those of $vec(\mathbf{H})$. From Theorem 7 and Corollary 7.1, $\Delta_M \to 0$ with the same rate as

$$\sum_{k=1}^M E\left(\left\|\mathbf{C}^{-1/2}\mathbf{x}_k\right\|^3\right).$$

Then from (55) and under the assumption that $\mathbf{C}$ does not depend on $M$,

$$\sum_{k=1}^M E\left(\left\|\mathbf{C}^{-1/2}\mathbf{x}_k\right\|^3\right)$$

converges to zero with at least the same rate as $|a_{1,M}| \to 0$.
∎

*C. Proof of Corollary 2.1:*

To prove the necessity of $1^{st}$ condition, let

$$\lim_{M \to \infty} \sum_{i=1}^k |a_{i,M}|^2 = c > 0.$$

Then,

$$|a_{1,M}| \geq \sqrt{c/k} > 0,$$

i.e. (9) is not satisfied. The sufficiency is trivial due to

$$\sum_{i=1}^k |a_{i,M}|^2 \geq |a_{1,M}|^2.$$

$2^{nd}$ condition follows from the following inequalities,

$$\begin{aligned}|a_{1,M}|^{2+\delta} &\leq \sum_{i=1}^M |a_{i,M}|^{2+\delta} \\ &\leq |a_{1,M}|^\delta \sum_{i=1}^M |a_{i,M}|^2 \\ &= |a_{1,M}|^\delta, \quad (57)\end{aligned}$$

∎

*D. Proof of Corollary 2.2:*

If $\mathbf{C}$ is non-singular, then $\lambda_k(\mathbf{C}) > 0$, $k = 1...n_t n_r$, where $\lambda_k(\mathbf{C})$ is the $k$-th eigenvalue of matrix $\mathbf{C}$. Without loss in generality assume that

$$\lambda_1(\mathbf{C}) \leq \lambda_2(\mathbf{C}) \leq ... \leq \lambda_{n_t n_r}(\mathbf{C}).$$

It is straightforward to show that

$$\begin{aligned}\lambda_1(\mathbf{C}) &\geq \sum_{k=1}^M |a_{k,M}|^2 \lambda_1(\mathbf{R}_{tk}^T \otimes \mathbf{R}_{rk}) \\ &= \sum_{k=1}^M |a_{k,M}|^2 \lambda_1(\mathbf{R}_{tk})\lambda_1(\mathbf{R}_{rk}) \quad (58)\end{aligned}$$

(i) If for every $k = 1...M$, $\lambda_1(\mathbf{R}_{tk}), \lambda_1(\mathbf{R}_{rk}) > 0$, i.e. all $\mathbf{R}_{tk}, \mathbf{R}_{tk}$ are non-singular, then $\lambda_k(\mathbf{C}) > 0$.

(ii) Let $S$ be a subset (either finite of infinite) of all singular $\mathbf{R}_{tk}, \mathbf{R}_{tk}$. Thus, if $\sum_{k \notin S} |a_{k,M}|^2 > 0$, from (58)

$$\lambda_1(\mathbf{C}) \geq \min_{k \notin S}\{\lambda_1(\mathbf{R}_{tk})\lambda_1(\mathbf{R}_{rk})\} \sum_{k \notin S} |a_{k,M}|^2 > 0 \quad (59)$$

∎

*E. Proof of Theorem 3:*

We start with the following theorem and lemmas.

*Theorem 8 (Martin and Ottersten [32]):* Let $C$ be instantaneous capacity (2) of a correlated Rayleigh-fading MIMO channel with Kronecker correlation structure, i.e. $\mathbf{H} \propto \mathbf{R}_r^{1/2}\mathbf{X}\mathbf{R}_t^{1/2}$, where $\mathbf{X}$ is an $n_r \times n_t$ i.i.d. Gaussian circular symmetric matrix, and $\mathbf{R}_t$ and $\mathbf{R}_r$ are normalized such that $n_t^{-1} tr\mathbf{R}_t = 1$ and $n_r^{-1} tr\mathbf{R}_r = 1$. (i) If

$$\lim_{n_t \to \infty} \frac{\|\mathbf{R}_t\|_2}{\|\mathbf{R}_t\|} = 0, \quad (60)$$

$C$ is asymptotically Gaussian in distribution as $n_t \to \infty$ with the mean $\mu$ and the variance $\sigma^2$ as follows

$$\mu = \ln \det\left(\mathbf{I} + \frac{\gamma_0}{n_r}\mathbf{R}_r\right), \quad (61)$$

$$\sigma^2 = \frac{1}{n_t^2} \|\mathbf{R}_t\|^2 \cdot \sum_{k=1}^{n_r} \left(\frac{\gamma_0 \lambda_k^r/n_r}{1 + \gamma_0 \lambda_k^r/n_r}\right)^2, \quad (62)$$

where $\lambda_k^r$, $k = 1...n_r$ are the eigenvalues of $\mathbf{R}_r$, and $\gamma_0$ is the total SNR at the Rx end.

(ii) Due to the symmetry in (5), the Theorem holds when Tx and Rx ends are exchanged.

As a side remark, we have the following additional result when both $n_t, n_r \to \infty$.

*Lemma 3:* Let $\mathbf{H}$ be an $n_r \times n_t$ matrix as in Theorem 8.
(i) if

$$\lim_{n_t,n_r \to \infty} n_r \|\mathbf{R}_t\|/n_t = 0,$$

then $\mathbf{H}\mathbf{H}^\dagger/n_t \overset{q.m.}{\to} \mathbf{R}_r$ as both $n_t, n_r \to \infty$, where $\overset{q.m.}{\to}$ denotes convergence in quadratic mean, i.e.

$$\lim_{n_t,n_r \to \infty} E \left\|\mathbf{H}\mathbf{H}^\dagger/n_t - \mathbf{R}_r\right\|^2 = 0,$$

from which $\mathbf{H}\mathbf{H}^\dagger/n_t \overset{p}{\to} \mathbf{R}_r$.
(ii) If

$$\lim_{n_t,n_r \to \infty} n_t \|\mathbf{R}_r\|/n_r = 0,$$

then $\mathbf{H}^\dagger\mathbf{H}/n_r \overset{q.m.}{\to} \mathbf{R}_t$ as both $n_t, n_r \to \infty$, from which $\mathbf{H}^\dagger\mathbf{H}/n_r \overset{p}{\to} \mathbf{R}_t$.

*Proof:* (i) First note that $E\{\mathbf{HH}^\dagger/n_t\} = \mathbf{R}_r$. Thus

$$E\left\|\mathbf{HH}^\dagger/n_t - \mathbf{R}_r\right\|^2 = n_t^{-2} tr\left(E\{\mathbf{HH}^\dagger\mathbf{HH}^\dagger\}\right) - \|\mathbf{R}_r\|^2 \quad (63)$$

Consider the trace in (63)

$$tr\left(E\{\mathbf{HH}^\dagger\mathbf{HH}^\dagger\}\right) = tr\left(E\{\mathbf{H}^\dagger\mathbf{HH}^\dagger\mathbf{H}\}\right)$$
$$= \sum_{m=1}^{n_t}\sum_{k=1}^{n_t} E\{\mathbf{h}_m^\dagger \mathbf{h}_k \mathbf{h}_k^\dagger \mathbf{h}_m\} \quad (64)$$
$$= \sum_{m=1}^{n_t}\sum_{k=1}^{n_t}\sum_{i=1}^{n_r}\sum_{j=1}^{n_r} E\{H_{im}^* H_{ik} H_{jk}^* H_{jm}\},$$

where $\mathbf{h}_k$ is the $k$-th column of $\mathbf{H}$, and $H_{ik}$ is an $i,k$-th element of $\mathbf{H}$. Since $H_{ik}$, $k=1...n_t$, $i=1...n_r$ are Gaussian circular symmetric, their fourth order cumulant is zero [54], i.e.

$$\sum_{m=1}^{n_t}\sum_{k=1}^{n_t}\sum_{i=1}^{n_r}\sum_{j=1}^{n_r} E\{H_{im}^* H_{ik} H_{jk}^* H_{jm}\}$$
$$= \sum_{m=1}^{n_t}\sum_{k=1}^{n_t}\sum_{i=1}^{n_r}\sum_{j=1}^{n_r} E\{H_{im}^* H_{ik}\} E\{H_{jk}^* H_{jm}\}$$
$$+ \sum_{m=1}^{n_t}\sum_{k=1}^{n_t}\sum_{i=1}^{n_r}\sum_{j=1}^{n_r} E\{H_{ik} H_{jk}^*\} E\{H_{im}^* H_{jm}\} \quad (65)$$

Following the Kronecker correlation model (13), $E\{H_{ij} H_{km}^*\} = T_{jm}^* R_{ik}$, where $T_{jm}$ and $R_{ik}$ are $j,m$-th and $i,k$-th elements of $\mathbf{R}_t$ and $\mathbf{R}_r$ respectively. Thus

$$\sum_{m=1}^{n_t}\sum_{k=1}^{n_t}\sum_{i=1}^{n_r}\sum_{j=1}^{n_r} E\{H_{im}^* H_{ik} H_{jk}^* H_{jm}\}$$
$$= \sum_{m=1}^{n_t}\sum_{k=1}^{n_t}\sum_{i=1}^{n_r}\sum_{j=1}^{n_r} \left(T_{km}^* R_{ii}^* T_{km} R_{jj} + T_{kk}^* R_{ij} T_{mm} R_{ij}^*\right)$$
$$= \sum_{m=1}^{n_t}\sum_{k=1}^{n_t} \left(n_r^2 |T_{km}|^2 + T_{kk}^* T_{mm} \|\mathbf{R}_r\|^2\right)$$
$$= n_r^2 \|\mathbf{R}_t\|^2 + n_t^2 \|\mathbf{R}_r\|^2 \quad (66)$$

Substituting (66) in (63), one obtains

$$E\left\|\mathbf{HH}^\dagger/n_t - \mathbf{R}_r\right\|^2 = \left(\frac{n_r}{n_t}\|\mathbf{R}_t\|\right)^2, \quad (67)$$

i.e. if $\lim_{n_t,n_r \to \infty} n_r \|\mathbf{R}_t\|/n_t = 0$, then

$$\lim_{n_t,n_r \to \infty} E\left\|\mathbf{HH}^\dagger/n_t - \mathbf{R}_r\right\|^2 = 0 \quad (68)$$

(ii) A proof is the same as above due to the symmetry of the problem. ∎

(i) The proof is based on three claims: 1) Under condition (14), a FRMK channel is asymptotically Rayleigh-fading in distribution, 2) Under condition (15), the capacity distribution of the FRMK channel is asymptotically Gaussian with the mean and variance (61), (62) respectively, and 3) At low SNR regime, the moments of the asymptotic Gaussian distribution are given by (18) and (19).

*Claim 1*: Consider the last inequality in (55) under the assumption that $n_t^{-2} E\|\mathbf{h}_{tk}\|^4 < \infty$ and $n_r^{-2} E\|\mathbf{h}_{rk}\|^4 < \infty$, $\mathbf{R}_t = \mathbf{R}_{tk}$, $\mathbf{R}_r = \mathbf{R}_{rk}$, $\forall k$, and $\mathbf{R}_t$, $\mathbf{R}_r$ are positive definite and normalized, so that $n_t^{-1} tr\mathbf{R}_t = 1$ and $n_r^{-1} tr\mathbf{R}_r = 1$:

$$\left\|\mathbf{C}^{-1/2}\right\|^3 \cdot \max_k \left(E\|\mathbf{h}_{tk}\|^4 \cdot E\|\mathbf{h}_{rk}\|^4\right)^{3/4} |a_{1,M}|$$
$$\leq \left\|(\mathbf{R}_t^T \otimes \mathbf{R}_r)^{-1/2}\right\|^3 M_0^{3/4} n_t^{3/2} n_r^{3/2} |a_{1,M}|$$
$$= \left\|\mathbf{R}_t^{-T/2} \otimes \mathbf{R}_r^{-1/2}\right\|^3 M_0^{3/4} n_t^{3/2} n_r^{3/2} |a_{1,M}|$$
$$= \left\|\mathbf{R}_t^{-1/2}\right\|^3 \left\|\mathbf{R}_r^{-1/2}\right\|^3 M_0^{3/4} n_t^{3/2} n_r^{3/2} |a_{1,M}|, \quad (69)$$

where $M_0$ is a finite number, such that

$$M_0 \geq \max_k \left(n_t^{-2} E\|\mathbf{h}_{tk}\|^4 \cdot n_r^{-2} E\|\mathbf{h}_{rk}\|^4\right).$$

Thus, if for any $n_t$,

$$\lim_{M \to \infty} n_t^{3/2} \left\|\mathbf{R}_t^{-1/2}\right\|^3 |a_{1,M}| = 0,$$

Theorem 2 applies, i.e. the FRMK channel is asymptotically Rayleigh-fading in distribution.

*Claim 2*: Following Theorem 8, the capacity distribution of the above FRMK channels is asymptotically Gaussian under condition (15) of Theorem 3 as $n_t \to \infty$. Consequently, the mean and the variance of the asymptotic capacity are given by (61) and (62) respectively.

*Claim 3*: Consider $\lambda_k^r/n_r$, $k=1...n_r$, where $\lambda_k^r$ are the eigenvalues of $\mathbf{R}_r$ and $\gamma_0 \lambda_k^r/n_r \ll 1$. Using the Maclaurin series of the right side hand of (61), one obtains

$$\mu = \sum_{k=1}^{n_r} \ln\left(1 + \frac{\gamma_0}{n_r} \cdot \lambda_k^r\right) \quad (70)$$
$$= \frac{\gamma_0}{n_r}\sum_{k=1}^{n_r} \lambda_k^r - \frac{1}{2}\left(\frac{\gamma_0}{n_r}\right)^2 \sum_{k=1}^{n_r} (\lambda_k^r)^2 + ...$$

which, under the normalization $n_r^{-1}\sum_{k=1}^{n_r} \lambda_k^r = 1$, yields $\mu \approx \gamma_0$. This proves (18). Applying the same approach on the right side hand of (62), one obtains

$$\sigma^2 = \frac{1}{n_t^2} \|\mathbf{R}_t\|^2 \cdot \sum_{k=1}^{n_r} \left(\frac{\gamma_0 \lambda_k^r/n_r}{1+\gamma_0 \lambda_k^r/n_r}\right)^2 \quad (71)$$
$$= \frac{1}{n_t^2} \|\mathbf{R}_t\|^2 \cdot \sum_{k=1}^{n_r} \left[(\gamma_0 \lambda_k^r/n_r)^2 - 2(\gamma_0 \lambda_k^r/n_r)^3 + ...\right]$$

i.e.

$$\sigma^2 \approx \gamma_0^2 \frac{1}{n_t^2}\|\mathbf{R}_t\|^2 \frac{1}{n_r^2}\|\mathbf{R}_r\|^2, \quad (72)$$

which proves (19).

(ii) A proof of part (ii) is the same due to the symmetry of the problem. ∎

*F. Proof of Theorem 5:*

Let $f$ be a scalar function defined on $\mathbf{R} \in \mathcal{M}_M$. $f$ is called Schur-convex if for any $\mathbf{R}_1, \mathbf{R}_2 \in \mathcal{M}_M$ such that $\mathbf{R}_1 \succ \mathbf{R}_2$, $f(\mathbf{R}_1) \geq f(\mathbf{R}_2)$ [37]. From [[55], Theorem 3.A.4], $f$ is Schur-convex iff

$$(\lambda_i - \lambda_j)\left(\frac{\partial f(\mathbf{R})}{\partial \lambda_i} - \frac{\partial f(\mathbf{R})}{\partial \lambda_j}\right) \geq 0, \quad (73)$$

where $\lambda_i$, $i = 1...n$ are the eigenvalues of $\mathbf{R}$. Let $f(\mathbf{R}) = \|\mathbf{R}\|^2$, then

$$(\lambda_i - \lambda_j)\left(\frac{\partial \|\mathbf{R}\|^2}{\partial \lambda_i} - \frac{\partial \|\mathbf{R}\|^2}{\partial \lambda_j}\right)$$
$$= (\lambda_i - \lambda_j)\left(\frac{\partial}{\partial \lambda_i}\sum_{k=1}^n \lambda_k^2 - \frac{\partial}{\partial \lambda_j}\sum_{k=1}^n \lambda_k^2\right)$$
$$= 2(\lambda_i - \lambda_j)^2 \geq 0, \qquad (74)$$

i.e. $\|\ \|^2$ is Schur-convex. Therefore, $\mathbf{R}_1 \succ \mathbf{R}_2$ iff $\|\mathbf{R}_1\| \geq \|\mathbf{R}_2\|$. ■

## REFERENCES

[1] I. E. Telatar, "Capacity of Multi-Antenna Gaussian Channels", *AT&T Bell Labs, Internal Tech. Memo*, June 1995, (*European Trans. Telecom.*, v.10, no. 6, Dec. 1999).

[2] G. J. Foschini and M. J. Gans, "On Limits of Wireless Communications in a Fading Environment when Using Multiple Antennas", *Wireless Personal Commun.*, vol. 6, no. 3, pp. 311-335, March 1998.

[3] S. H. Simon and A. L. Moustakas, "Optimizing MIMO Antenna Systems with Channel Covariance Feedback", *IEEE Jour. on Sel. Areas Commun.*, vol. 21, no. 3, pp. 406-417, Apr. 2003.

[4] M. Chiani, M. Z. Win and A. Zanella, "On the Capacity of Spatially Correlated MIMO Rayleigh-Fading Channels", *IEEE Trans. on Information Theory*, vol. 49, no. 10, pp. 2363-2371, Oct. 2003.

[5] H. Shin, M. Z. Win, J. H. Lee and M. Chiani, "On the Capacity of Doubly Correlated MIMO Channels", *IEEE Trans. on Wireless Commun.*, vol. 5, no. 8, pp. 2253-2265, Aug. 2006.

[6] M. Kang and M. S. Alouini, "Capacity of MIMO Rician Channels", *IEEE Trans. on Wireless Commun.*, vol. 5, no. 1, pp. 112-122, Jan. 2006.

[7] M. R. McKay and I. B. Collings, "General Capacity Bounds for Spatially Correlated Rician MIMO Channels", *IEEE Trans. on Information Theory*, vol. 51, no.9, pp. 3121-3145, Sep. 2005.

[8] G. Fraidenraich, O. Leveque, and J. M. Cioffi, "On the MIMO Channel Capacity for the Nakagami-m Channel", *IEEE Transactions on Information Theory*, vol. 54, no. 8, pp. 3752-3757, Aug. 2008.

[9] M. Dohler, and H. Aghvami, "Information Outage Probability of Distributed STBCs over Nakagami Fading Channels," *IEEE Communications Letters*, vol. 8, no. 7, pp. 437-439, July 2004.

[10] D. Chizhik, G. J. Foschini and R. A. Valenzuela, "Capacities of Multi-Element Transmit and Receive Antennas: Correlations and Keyholes", *Electronics Letters*, vol. 36, no. 13, pp. 1099-1100, June 2000.

[11] D. Chizhik, G. J. Foschini, M. J. Gans, and R. A. Valenzuela, "Keyholes, Correlations, and Capacities of Multielement Transmit and Receive Antennas," *IEEE Transactions on Wireless Communications*, vol. 1, no. 2, pp. 361-368, Apr. 2002.

[12] S. M. Alamouti, "A Simple Transmit Diversity Technique for Wireless Communications ," *IEEE Journal on Selected Areas in Communications*, vol. 16, no. 8, pp. 1451-1458, Oct 1998.

[13] D. Gesbert, H. Bolcskei, D. A. Gore, A. J. Paulraj, "Outdoor MIMO Wireless Channels: Models and Performance Prediction," *IEEE Transactions on Communications*, vol. 50, no. 12, pp. 1926-1934, Dec 2002.

[14] H. Xu, M. J. Gans, N. Amitay, and R. A. Valenzuela, "Experimental Verification of MTMR System Capacity in Controlled Propagation Environment", *Electronics Letters*, vol. 37, no. 15, pp. 936-937, July 2001.

[15] D. Porrat, P. Kyritsi, and D. C. Cox, "MIMO Capacity in Hallways and Adjacent Rooms", *IEEE Globecom '02*, vol. 2, pp.17-21, Taipei, Taiwan, Nov. 2002.

[16] P. Almers, F. Tufvensson and A. F. Molisch, "Measurement of Keyhole Effect in a Wireless Multiple-Input Multiple-Output (MIMO) Channel", *IEEE Communications Letters*, vol. 7, no. 8, pp. 373-375, Aug. 2003.

[17] P. Almers, F. Tufvensson and A. F. Molisch, "Keyhole Effect in MIMO Wireless Channels: Measurements and Theory," *IEEE Transactions on Wireless Communications*, vol. 5, no. 12, pp. 3596-3604, December 2006.

[18] S. Loyka, "Multiantenna Capacities of Waveguide and Cavity Channels," *IEEE Transactions on Vehicular Technology*, vol. 54, no. 3, pp. 863-872, May 2005.

[19] G. Levin and S. Loyka, "Multi-Keyholes and Measure of Correlation in MIMO Channels", *in Proc. QBSC'06, $23^{rd}$ Biennial Symposium on Communications*, Kingston, ON, May-June 2006.

[20] G. Levin and S. Loyka, "Multi-Keyhole MIMO Channels: Asymptotic Analysis of Outage Capacity", *in Proc. IEEE 2006 ISIT, 2006 IEEE International Symposium on Information Theory*, Seattle, WA, July 2006.

[21] G. Levin and S. Loyka, "On the Outage Capacity Distribution of Correlated Keyhole MIMO Channels", *IEEE Trans. on Information Theory*, vol. 54, no. 7, pp. 3232-3245, July 2008.

[22] H. Shin and J. H. Lee, "Capacity of Multiple-Antenna Fading Channels: Spatial Fading Correlation, Double Scattering, and Keyhole", *IEEE Trans. on Information Theory*, vol. 49, no. 10, pp. 2636-2646, Oct. 2003.

[23] X. W. Cui and Z. M. Feng, "Lower Capacity Bound for MIMO Correlated Fading Channels with Keyhole", *IEEE Communications Letters*, vol. 8, no. 8, pp. 500-5002, Aug. 2004.

[24] S. Sanayei, A. Hedayat and A. Nosratinia, "Space Time Codes in Keyhole Channels: Analysis and Design," *IEEE Transactions on Wireless Communication* , vol. 6, no. 6, pp. 2006-2011, June 2007.

[25] S. Sanayei, and A. Nosratinia, "Antenna Selection in Keyhole Channels," *IEEE Transactions on Communications*, vol. 55, no. 3, pp. 404-408, March 2007.

[26] H. Shin, aqnd J. H. Lee, "Performance Analysis of Space-Time Block Codes over Keyhole Nakagami-m Fading Channels," *IEEE Transactions on Vehicular Technology*, vol.53, no.2, pp. 351-362, March 2004.

[27] A. M. Sayeed, "Deconstructing multiantenna fading channels," *IEEE Transactions on Signal Processing*, vol.50, no.10, pp. 2563-2579, Oct 2002.

[28] Special Issue on Models, Theory, and Codes for Relaying and Cooperation in Communication Networks, *IEEE Transactions on Information Theory*, vol.53, no. 10, Oct. 2007.

[29] J. P. Kermoal, L. Schumacher, K. I. Pedersen, P.E. Mogensen and F. Frederiksen, "A Stochastic MIMO Radio Channel Model with Experimental Validation", *IEEE Jour. on Sel. Areas Commun.*, vol. 20, no. 6, pp. 1211-1226, Aug. 2002.

[30] B. M. Hochwald, T. L. Marzetta and V. Tarokh, "Multiple-Antenna Channel Hardening and Its Implications for Rate Feedback and Scheduling", *IEEE Trans. on Information Theory*, vol. 50, no. 9, pp. 1893-1909, Sept. 2004.

[31] A. L. Moustakas, S. H. Simon and A. M. Sengupta, "MIMO Capacity Through Correlated Channels in the Presence of Correlated Inteferers and Noise: A (Not So) Large N Analysis", *IEEE Trans. on Information Theory*, vol. 49, no. 10, pp. 2545-2561, Oct. 2003.

[32] C. Martin and B. Ottersten, "Asymptotic Eigenvalue Distributions and Capacity for MIMO Channels under Correlated Fading", *IEEE Trans. on Wireless Commun.*, vol. 3, no. 4, pp. 1350-1358, July 2004.

[33] A. M. Tulino and S. Verdu, "Random Matrix Theory and Wireless Communications", *Foundations and Trends in Commun. and Inform. Theory*, vol. 1, pp. 1-182, 2004.

[34] A. Lozano, A. M. Tulino, and S. Verdu, "Multiple-Antenna Capacity in the Low-Power Regime", *IEEE Transactions on Information Theory*, vol. 49, no. 10, pp. 2527- 2544, Oct. 2003.

[35] M. T. Ivrlac and J. A. Nossek, "Diversity and Correlation in Rayleigh Fading MIMO Channels", *in Proc. of 2005 IEEE $61^{st}$ Vehicular Technology Conference, VTC2005-Spring*, vol. 1, pp. 151-155, Stockholm, Sweden, May-June 2005.

[36] H. Boche, E. A. Jorswieck, "Outage Probability of Multiple Antenna Systems: Optimal Transmission and Impact of Correlation," , *2004 International Zurich Seminar on Communications*, pp. 116-119, Switzerland, Feb. 2004.

[37] H. Bosche and E. A. Jorswieck, "On the Ergodic Capacity as a Function of the Correlation Properties in Systems with Multiple Transmit Antennas without CSI at the Transmitter", *IEEE Trans. on Commun.*, vol. 52, no. 10, pp. 1654-1657, Oct. 2004.

[38] A. M. Tulino, A. Lozano, and S. Verdu, "Impact of Antenna Correlation on the Capacity of Multiantenna Channels," *IEEE Transactions on Information Theory*, vol. 51, no. 7, pp. 2491-2509, July 2005.

[39] S. Loyka and G. Levin, "On Physically–Based Normalization of MIMO Channel Matrix", *IEEE Trans. on Wireless Commun.*, vol.8, no.3, pp.1107-1112, March 2009.

[40] S. Loyka, "Channel Capacity of MIMO Architecture Using the Exponential Correlation Matrix", *IEEE Communication Letters*, vol. 5, no. 9, pp. 369-371, Sep. 2001.

[41] L. Zheng, and D. N. C. Tse, "Diversity and Multiplexing: A Fundamental Tradeoff in Multiple-Antenna Channels," *IEEE Transactions on Information Theory*, vol.49, no.5, pp. 1073-1096, May 2003.

[42] T. S. Ferguson, *A Course in Large Sample Theory*, Chapman & Hall/CRC, $1^{st}$ Ed. Reprint, 2002.

[43] D. S. Bernstein, *Matrix Mathematics: Theory, Facts, and Formulas with Applications to Linear Systems Theory*, Princeton University Press, 2005.

[44] G. Levin and S. Loyka, "Comments on Asymptotic Eigenvalue Distributions and Capacity for MIMO Channels under Correlated Fading", *IEEE Trans. on Wireless Communications*, vol. 7, no. 2, pp. 475-479, Feb. 2008.

[45] H. Shin, and M. Z. Win, "MIMO Diversity in the Presence of Double Scattering," *IEEE Trans. Inform. Theory*, revised for publication. [Online]. Available: http://arxiv.org/abs/cs/0511028.

[46] M. K. Simon and M. S. Alouini, *Digital Communication over Fading Channels: A Unified Approach to Performance Analysis*, John Wiley & Sons, Inc., 2000.

[47] T. S. Chu and L. J. Greenstein, "A Semi-Empirical Representation of Antenna Diversity Gain at Cellular and PCS Base Stations," *IEEE Transactions on Communications*, , vol. 45, no. 6, pp. 644-646, Jun 1997.

[48] S. Loyka, J. Mosig, "Channel Capacity of N-Antenna BLAST Architecture", *Electronics Letters*, vol. 36, no.7, pp. 660-661, Mar. 2000.

[49] E. A. Jorswieck and H. Boche, "Outage Probability in Multiple Antenna Systems," *Euro. Trans. Telecomms,* vol. 18, no. 5, pp. 445-456, Apr. 2007.

[50] A. M. Mathai, S. B. Provost, *Quadratic Forms in Random Variables*, Marcel Dekker, Inc., New York, Basel, Hong Kong, 1992.

[51] G. Levin, *Capacity Analysis of Asymptotically Large MIMO Channels*, Ph.D. Thesis, University of Ottawa, ON, Canada, 2008.

[52] V. Bentkus, "A Lyapunov-Type Bound in $\Re^d$", *Theory Probab. Appl.*, vol. 49, no. 2, pp. 311-323, 2005.

[53] M. Fisz, *Probability Theory and Mathematical Statistics*, John Willey & Son, Inc., $3^{rd}$ Ed., 1963.

[54] B. Porat, *Digital Processing of Random Signals: Theory and Methods*, Prentice-Hall, Inc., Englewood Cliffs, New Jersey, 1994.

[55] A. W. Marshall and I. Olkin, *Inequalities: Theory of Majorization and Its Applications*, vol. 143, Mathematics in Science and Engineering, London, U.K. Academic, 1979.

[56] S. Loyka, G. Levin, Finite-SNR Diversity-Multiplexing Tradeoff via Asymptotic Analysis of Large MIMO Systems, *IEEE Transactions on Information Theory,* vol. 56, no. 10, pp. 4781-4792, Oct. 2010.

[57] J. Von Neumann, Some matrix-inequalities and metrization of matric-space, *Tomsk Univ. Rev.* 1 (1937) 286-300.

[58] S. Jin, M. R. McKay, K-K. Wong, X. Gao, "Transmit Beamforming in Rayleigh Product MIMO Channels: Capacity and Performance Analysis", *IEEE Trans. Sig. Proc.*, vol. 56, no. 10, pp. 5204-5221, Oct. 2008.

[59] C. Zhong. S. Jin, K-K. Wong, M. R. McKay, "Outage Analysis for Optimal Beamforming MIMO Systems in Multikeyhole Channels", *IEEE Trans. Sig. Proc.*, vol. 58, no. 3, pp. 1451-1456, Mar. 2010.

[60] C. Zhong. S. Jin, K-K. Wong, M. R. McKay, "Performance Analysis of Optimal Joint Beamforming in Multi-Keyhole MIMO Channels", *IEEE International Conference on Communications (ICC 2009),* June 14-18, Dresden, Germany.

[61] G. Levin, S Loyka, "Diversity-Multiplexing Tradeoff and Outage Probability in MIMO Relay Channels," *2010 IEEE International Symposium on Information Theory Proceedings (ISIT 2010),* pp. 2218-2222, June 13-18, 2010, Austin, TX.

[62] D. N. C. Tse, P. Viswanath, *Fundamentals of Wireless Communications*, Cambridge University Press, 2005.

[63] W. L. Root, P. P. Varaya, "Capacity of Classes of Gaussian Channels", *SIAM J. Appl. Math.*, vol. 16, no. 6, pp. 1350-1393, Nov. 1968.

[64] S. Verdu, T.S. Han, "A General Formula for Channel Capacity", *IEEE Transactions on Information Theory*, vol. 40, no. 4, pp. 1147-1157, July 1994.

[65] M. Effros, A. Goldsmith, Y. Liang, "Generalizing Capacity: New Definitions and Capacity Theorems for Composite Channels," *IEEE Transactions on Information Theory*, vol. 56, no. 7, pp. 3069-3087, July 2010.

[66] D. P. Palomar, J. M. Cioffi, M. A. Lagunas, "Uniform Power Allocation in MIMO Channels: A Game-Theoretic Approach," *IEEE Transactions on Information Theory*, vol. 49, no. 7, pp. 1707-1727, July 2003.

**Georgy Levin** received the B.S. and M.S. degrees, both cum laude, in Electrical and Computer Engineering from Ben-Gurion University of the Negev, Israel in 1995 and 2000, and the Ph.D. degree from the University of Ottawa, Ontario, Canada in 2008. He is currently a research assistant at the University of Ottawa. His research spans the fields of wireless communications and information theory with specific interest in MIMO systems, smart antennas, relay networks and cognitive radio. Dr. Levin is a reviewer for numerous IEEE periodicals and conferences. He received a number of awards from Canada, Israel and former USSR governments.

**Sergey Loyka** (M'96–SM'04) was born in Minsk, Belarus. He received the Ph.D. degree in Radio Engineering from the Belorussian State University of Informatics and Radioelectronics (BSUIR), Minsk, Belarus in 1995 and the M.S. degree with honors from Minsk Radioengineering Institute, Minsk, Belarus in 1992. Since 2001 he has been a faculty member at the School of Information Technology and Engineering, University of Ottawa, Canada. Prior to that, he was a research fellow in the Laboratory of Communications and Integrated Microelectronics (LACIME) of Ecole de Technologie Superieure, Montreal, Canada; a senior scientist at the Electromagnetic Compatibility Laboratory of BSUIR, Belarus; an invited scientist at the Laboratory of Electromagnetism and Acoustic (LEMA), Swiss Federal Institute of Technology, Lausanne, Switzerland. His research areas include wireless communications and networks, MIMO systems and smart antennas, RF system modeling and simulation, and electromagnetic compatibility, in which he has published extensively. Dr. Loyka is a technical program committee member of several IEEE conferences and a reviewer for numerous IEEE periodicals and conferences. He received a number of awards from the URSI, the IEEE, the Swiss, Belarus and former USSR governments, and the Soros Foundation.